\documentclass[%
reprint,
superscriptaddress,
 amsmath,
 amssymb,
 aps,
 onecolumn,
]{revtex4-2}
\usepackage{graphicx}% Include figure files
\usepackage{dcolumn}% Align table columns on decimal point
\usepackage{bm}% bold math
\usepackage{epstopdf}
\usepackage{xcolor}
\newcommand{\review}[1]{{\color{black}#1}}
\newcommand{\reviewtwo}[1]{{\color{blue}#1}}
\begin{document}

%\preprint{APS/123-QED}

\title{Taylor dispersion of elongated rods}

\author{Ajay Harishankar Kumar}
\affiliation{Brown University, Center for Fluid Mechanics and School of Engineering, 184 Hope St., Providence RI 02912.}
\author{Stuart J. Thomson}
\affiliation{Brown University, Center for Fluid Mechanics and School of Engineering, 184 Hope St., Providence RI 02912.}
\author{Thomas R. Powers}
\affiliation{Brown University, Center for Fluid Mechanics and School of Engineering, 184 Hope St., Providence RI 02912.}
\affiliation{Brown University, Brown Theoretical Physics Center and Department of Physics, 184 Hope St., Providence RI 02912.}
\author{Daniel M. Harris}
\email[]{daniel\_harris3@brown.edu}
\affiliation{Brown University, Center for Fluid Mechanics and School of Engineering, 184 Hope St., Providence RI 02912.}

\date{\today}

\begin{abstract}%working abstract
Particles transported in fluid flows, such as cells, polymers, or nanorods, are rarely spherical. In this study, we numerically and theoretically investigate the dispersion of an initially localized patch of passive elongated Brownian particles in a two-dimensional Poiseuille flow, demonstrating that elongated particles exhibit an enhanced longitudinal dispersion. In a shear flow, the rods translate due to advection and diffusion and rotate due to rotational diffusion and their classical Jeffery’s orbit. The magnitude of the enhanced dispersion depends on the particle's aspect ratio and the relative importance of its shear-induced rotational advection and rotational diffusivity. When rotational diffusion dominates, we recover the classical Taylor dispersion result for the longitudinal spreading rate using an orientationally averaged translational diffusivity for the rods. However, in the high-shear limit, the rods tend to align with the flow and ultimately disperse more due to their anisotropic diffusivities. Results from our Monte Carlo simulations of the particle dispersion are captured remarkably well by a simple theory inspired by Taylor's original work. For long times and large Peclet numbers, an effective one-dimensional transport equation is derived with integral expressions for the particles' longitudinal transport speed and dispersion coefficient. The enhanced dispersion coefficient can be collapsed along a single curve for particles of high aspect ratio, representing a simple correction factor that extends Taylor's original prediction to elongated particles.

\end{abstract}

\maketitle

\section{Introduction}
Understanding the transport of particles in fluid flow has led to the development of novel particle separation techniques, mixing strategies, and lab-on-a-chip devices ~\cite{squires2005microfluidics,stone2004engineering}. In many practical cases of interest, the geometry of the particles themselves may be complex \cite{witten2020review}, and hence it is important to understand how their shape \cite{truong2015importance} influences their bulk transport. Herein, we study how the elongated shape of passive, rod-like Brownian particles affects their dispersion in a steady, two-dimensional Poiseuille flow.

\begin{figure}[ht]
    \centering
    \includegraphics{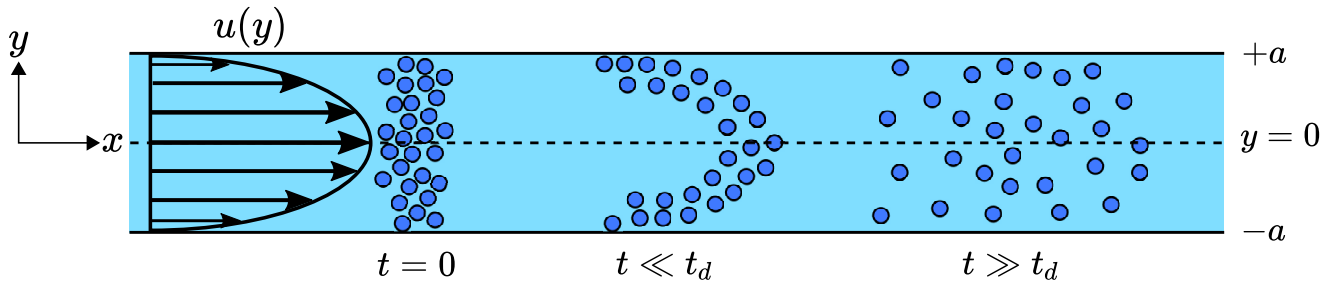}
    \caption{Illustration of the classical Taylor dispersion process. At early times ($t\ll t_d$) a plug of non-interacting Brownian tracer particles mimics the shape of the flow. The shear flow induces lateral concentration gradients that molecular diffusion tends to minimize.  The overall effect at late times ($t\gg t_d$) is an enhanced diffusive-like longitudinal spreading of particles as the solute patch is advected downstream at the mean speed of the fluid flow.}
    \label{fig:Dispersion}
\end{figure}

In a seminal paper \cite{taylor1953dispersion}, Taylor quantified the dispersion of spherical solute particles subject to Poiseuille flow in a cylindrical pipe. In Taylor's original physical picture (see Figure \ref{fig:Dispersion}), when a uniform patch of a solute is injected in a laminar flow, it spreads due to the combined effects of advection and diffusion. At early times, the solute patch mimics the shape of the parabolic flow profile, inducing lateral concentration gradients that drive net lateral transport by molecular diffusion. Ultimately, the shear flow enhances the spreading of the solute, a phenomenon now known as Taylor dispersion. Later, Aris expanded on Taylor's results in more rigorous mathematical detail using the method of moments, and thus this phenomenon is also frequently referred to as Taylor-Aris dispersion \cite{aris1956dispersion}. Perhaps the most complete mathematical treatment is due to Frankel \& Brenner \cite{frankel1989foundations}, who derived a generalized theory of Taylor-Aris dispersion. This robust framework has since been used to solve a wide class of dispersion problems, including the dispersion of active matter in shear flow \cite{hill2002taylor,manela2003generalized,jiang2019dispersion,jiang2020dispersion}.
Of most relevance to the present work, Peng and Brady studied the upstream swimming and dispersion of active Brownian particles in a two-dimensional Poiseuille flow with one degree of rotational freedom for spherical and rod-shaped particles, demonstrating enhancement of the dispersion factor for active Brownian particles due to their swimming (i.e.\ activity) \cite{peng2020upstream}. Such an enhancement was observed experimentally for bacteria in porous media \cite{dehkharghani2019bacterial}. Elsewhere, the effect of channel geometry on the dispersion of passive tracers has been well-documented to control or enhance the dispersion properties \cite{dutta2006effect,aminian2016boundaries,bernardi2018space,lee2021dispersion}, while the effect of the dispersion factor on pulsatile flow has also been documented ~\cite{marbach2019active,salerno2020aris}.
Previous studies have also focused on the Brownian motion of ellipsoidal \cite{han2006brownian,han2009quasi} and boomerang-shaped particles \cite{chakrabarty2013brownian} in the absence of external flow. However, despite these advances, the effect of a passive particle's shape on dispersion in the presence of flow has received relatively little attention. 

It is now well-known that confined rod-shaped particles or fibres have a tendency to migrate towards channel walls when subject to a background shear flow \cite{agarwal1994migration,schiek1997cross,jendrejack2004shear,makino2005migration,fu2009separation}. This effect was characterized by Nitsche \& Hinch \cite{nitsche1997shear}, who studied the lateral migration velocity and resultant distribution of rod-shaped particles in quasi-two-dimensional shear flow, assuming a uniform particle concentration in the longitudinal direction.  In complement to this prior work, we characterize the {\it longitudinal} transport properties of an initial concentration of confined Brownian rods in two-dimensional Poiseuille flow, using both Monte Carlo simulations and theoretical considerations. The rods are non-interacting Brownian tracers and modeled as elongated ellipsoids with the neglect of wall-based hydrodynamic effects. Our study reveals and quantifies two main results: a reduced mean transport speed for the rods compared to the mean speed of the fluid, and an enhanced rate of longitudinal dispersion compared to spherical particles. 

In the remainder of this section, we review Taylor's classical analysis applied to spherical particles in two-dimensional Poiseuille flow \cite{taylor1953dispersion}, followed by a discussion of extra physical considerations relevant for elongated particles. In \S\ref{sec:MC}, we describe our Monte Carlo method for calculating the dispersion coefficient for ellipsoidal particles in a two-dimensional Poiseuille flow. We then turn to a simplified theoretical analysis in the spirit of Taylor's original calculation in \S\ref{sec:simp_theory}, deriving semi-analytical expressions for the mean speed of the particles and the dispersion coefficient, in excellent agreement with the Monte Carlo simulations. We conclude with a summary of our results in \S\ref{conclusion}.

% Taylor showed that when a uniform patch of a solute is injected in a laminar flow, it spreads due to the combined effects of advection and diffusion as seem in Figure \ref{fig:Dispersion}. At early times, the solute patch mimics the shape of the flow profile, inducing lateral concentration gradients that drive net lateral transport by molecular diffusion. Ultimately, the shear enhances the effective dispersion of solute, a phenomenon commonly referred to as Taylor dispersion \cite{taylor1953dispersion}. We will review Taylor's results in brief.

Consider a parallel plate channel separated by a distance of $2 a$ with a fully developed Poiseuille flow with a maximum velocity of $U$ at $y = 0$, as depicted in Figure \ref{fig:Dispersion}. For isotropic solute particles with a characteristic diffusion constant $D$, a diffusive time scale can be defined as $t_d = a^2/D$, which is the characteristic time for a solute particle to travel from the center of the channel to the walls purely through molecular diffusion. There are two primary mechanisms of particle transport in this problem, advection and diffusion, the relative importance of which is characterized by the Peclet number
\begin{equation}\label{Pe_def_spheres}
    \mathrm{Pe} = \frac{U a}{D}.
\end{equation}
For long times, specifically $t \gg t_d$, and large $\mathrm{Pe}$ (advection dominated), Taylor characterized the laterally averaged concentration profile $\mathcal{C}_m(x,t)$ with an effective dispersion constant $\kappa_s$ that depends on the properties of the flow, channel geometry, and particles~\cite{taylor1953dispersion}. Taylor's original calculation was performed for a circular pipe, but the same analysis can be readily applied to describe dispersion in a two-dimensional channel (i.e.\ infinite parallel plates). Nondimensionalizing time using the diffusive time scale $t_d$, and the lengths $x$ and $y$ using the half-width $a$ of the channel, we find the dimensionless form of the laterally averaged transport equation to be the one-dimensional advective-diffusion equation
\begin{equation}\label{eqn:Taylor_result}
    \frac{\partial \mathcal{C}_m}{\partial t} + \frac{2}{3} \mathrm{Pe} \frac{\partial \mathcal{C}_m}{\partial x} =  \kappa_s \frac{\partial^2 \mathcal{C}_m}{\partial x^2},
\end{equation}
where the dimensionless effective dispersion constant is
\begin{equation}\label{kappa_s}
    \kappa_s = \frac{8}{945} \mathrm{Pe}^2.
\end{equation}
The dimensional effective dispersion constant $\kappa_s'$ is
\begin{equation}
    \kappa_s' = D \kappa_s = \frac{8}{945} \frac{U^2 a^2}{D}.
    \label{kappa-s-prime-eq}
\end{equation}

\review{As $\mathrm{Pe}\gg 1$, equations (\ref{eqn:Taylor_result}) and (\ref{kappa_s}) imply a significant increase in the longitudinal spreading rate resulting from the parallel shear flow. Relevant to more moderate Peclet numbers, Aris's rigorous expansion \cite{aris1956dispersion} introduced a correction to the expression of the effective dispersion constant, which accounts for the additional contribution due to the presence of molecular diffusion in the longitudinal direction: 
\begin{equation}
    \kappa_{s^{*}}' = D (\kappa_s + 1).    
\end{equation}
In the present study, we focus on the advection-dominated regime ($\text{Pe}\gg 1$), coinciding with that originally considered by Taylor for spherical particles.} 
We also note from equation \eqref{kappa-s-prime-eq} that the effective dispersion coefficient is {\it inversely} related to the molecular diffusion constant of the particle. In Taylor's analysis, the contribution of molecular diffusion to the expression for the effective dispersion, $\kappa_s$, arises exclusively from the lateral ($y$) diffusion term in the advection-diffusion equation governing the concentration of particles. Thus, in scenarios where the diffusion may be anisotropic (for example, when the solute particles are non-spherical, or when their diffusivity depends on $y$), the lateral diffusion coefficient, $D_y$, is the appropriate value to consider in such a scaling to estimate the effective dispersion constant. We will now discuss important quantities pertaining to ellipsoidal particles in a fluid.
\par
\par
\begin{figure}[ht]
    \centering
    \includegraphics{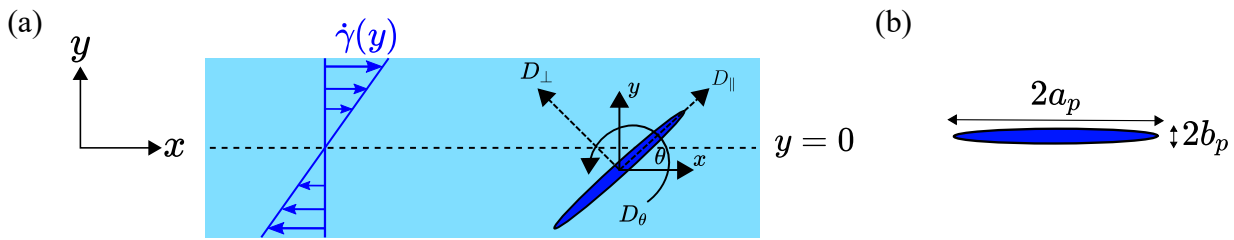}
    \caption{(a) Zoomed in schematic of a channel with a shear rate $\Dot{\gamma} (y)$. The figure also depicts the coordinate axes for each particle in the channel and its translational diffusivities along its perpendicular and parallel directions along with the rotational diffusivity. (b) %Shows 
    The definition of the semi-major axis $a_p$ and the semi-minor axis $b_p$.}
    \label{fig:coordinates}
\end{figure}
\begin{figure}
    \centering
    \includegraphics[width=1\textwidth]{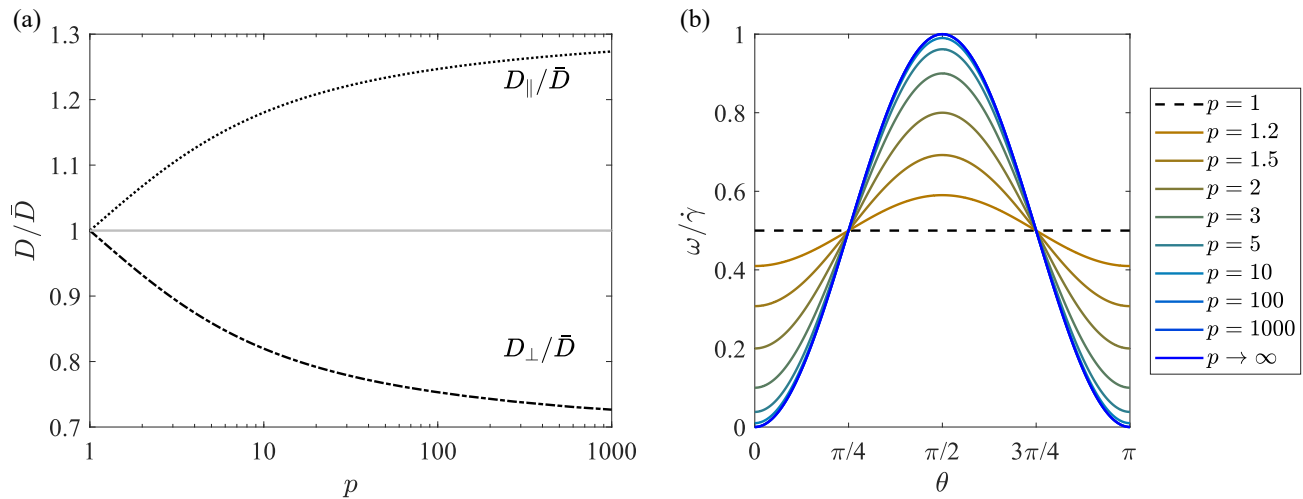}
    \caption{(a) Plot of $D_{\perp}{/\Bar{D}}$ (dash-dotted curve) and $D_{\parallel}/\Bar{D}$ (dotted curve). (b) The dimensionless rotation rate for different aspect ratios as a function of the angle $\theta$ between the rod axis and the flow direction.  In the absence of Brownian motion, the rods rotate the fastest when aligned normal to the direction of flow and rotate most slowly when aligned in the flow direction, spending more time in each revolution aligned with the flow.}
    \label{fig:curvesJO_D}
\end{figure}

The diffusion constants for an ellipsoidal particle constrained to translate and rotate in a plane follow from the Stokes-Einstein relation~\cite{berg1993random,han2006brownian,han2009quasi}.
Rotational and translational diffusion for an ellipsoidal particle are decoupled due to its symmetry~\cite{brenner1965coupling,brenner1967coupling,wegener1981diffusion}. 
The translational diffusion constants $D_\parallel$ and $D_\perp$ for a prolate ellipsoid are labeled in Figure~\ref{fig:coordinates}, and are given by~\cite{Perrin1936,happel2012low}
\begin{eqnarray}\label{eqDpara}
    D_{\parallel} &=& \frac{k_b T}{16 \pi \mu a_{p}} p \left[-\frac{2 p}{p^2 - 1} + \frac{2 p^2 - 1}{\left(p^2 -1 \right)^{3/2}} \log\left({\frac{p + \sqrt{p^2 - 1}}{p - \sqrt{p^2 - 1}}} \right)   \right],\\ \label{eqDperp}
    D_{\perp} &=& \frac{k_b T} {16 \pi \mu a_{p}} p \left[\frac{p}{p^2 - 1} + \frac{2 p^2 - 3}{\left(p^2 -1 \right)^{3/2}} \log\left({p + \sqrt{p^2 - 1}} \right)   \right],
\end{eqnarray}
where $k_b$ is Boltzmann's constant, $T$ is temperature, $p\equiv a_p/b_p$ is the ratio of the semi-major and semi-minor axes of the particle, and $\mu$ is the dynamic viscosity. Note that for prolate ellipsoids, $p>1$, 
and $D_\parallel\rightarrow 2D_\perp$ in the ``slender-body" limit $p\rightarrow\infty$.
We define an orientationally averaged diffusivity as
\begin{equation}\label{Dbar}
    \Bar{D}=\frac{D_{\perp} + D_{\parallel}}{2} .
\end{equation}
Figure \ref{fig:curvesJO_D}(a) %plots
shows how $D_\perp$ and $D_\parallel$ depend on the aspect ratio. A particle diffuses more readily along its long axis than against it.  The rotational diffusion constant is~\cite{Perrin1934,Koenig1975}
\begin{equation}\label{D_thetaeq}
    D_{\theta} = \frac{3 k_b T}{16 \pi \mu  a_p^3} \frac{p^4}{ p^4 - 1} \left[ \frac{\left(2 p^2 -1\right)\log\left({p + \sqrt{p^2 -1}}\right)}{p\sqrt{p^2 - 1}} -1 \right].
\end{equation}
We note that equations \eqref{eqDpara}, \eqref{eqDperp} and, \eqref{D_thetaeq} are commonly used to study the Brownian motion of ellipsoids confined to one degree of rotational freedom ~\cite{han2006brownian,han2009quasi}.

%%% Jeffery's orbit

Ellipsoidal particles rotate in a shear flow with a non-uniform rotational velocity in so-called Jeffery's orbits~\cite{jeffery1922motion}. For a prolate spheroid confined to one degree of rotational freedom in the plane within a two-dimensional Stokes flow, the rotation rate $\omega$ is a function of its angle $\theta$ relative to the flow \cite{bretherton1962motion}, specifically
\begin{equation}\label{JO}
    \omega(\theta) = \Dot{\gamma}  \ \frac{p^2 \sin^2{\theta} + \cos^2{\theta}}{p^2 + 1},
\end{equation}
where $\Dot{\gamma}$ is the local shear rate. In the slender-body limit ($p\rightarrow\infty$), the expression of the rotation rate reduces to $\omega(\theta) = \Dot{\gamma} \sin^2{\theta}$.  Equation (\ref{JO}) is plotted in Figure \ref{fig:curvesJO_D}(b), which shows that elongated particles ($p>1$) rotate fastest along the direction of the flow and rotate slowest normal to the direction of flow. Therefore, rods tend to spend more time aligned with the flow during a complete orbit. For a parabolic velocity profile, the shear rate is a linear function across the channel with the largest magnitude at the walls, as depicted in Figure \ref{fig:coordinates}. 
The rotational degree of freedom prompts us to define a rotational Peclet number
\begin{equation}\label{Rotational Peclet}
    \mathrm{Pe_r} = \frac{U}{a D_{\theta}}
\end{equation}
characterizing the ratio of the shear rate to rotational diffusion. For the case of a linear Couette shear flow, previous work has focused on describing how weak Brownian motion affects the three-dimensional Jeffery orbits~\cite{leal1971effect}. More recent work has explored the purely rotational analog of Taylor dispersion in which shear leads to a higher dispersion coefficient for rotation~\cite{Leahyetal2013,leahy2015effect}. As mentioned previously, for the case of a Poiseuille flow, ellipsoidal particles (unlike spherical particles) are known to migrate to the channel walls due to their anisotropic diffusivities and different alignments at different local shear rates \cite{fu2009separation,nitsche1997shear,makino2005migration,agarwal1994migration}.

%%%Monte Carlo
\section{Monte Carlo Simulation}
In this section, we model and simulate the dynamics of individual Brownian rods subject to a Poiseuille flow to deduce macroscopic statistical quantities, specifically the mean particle speed and dispersion coefficient using Monte Carlo simulation. The results show an enhanced dispersion for elongated particles and allow us to establish a simple physical picture for the phenomenon and its parametric dependencies.  \review{ The system is assumed to be in the dilute limit where particle-particle interactions are neglected. The system is assumed to be in the dilute limit where particle-particle interactions are neglected. We note that this assumption becomes more accurate as time progresses and the solute disperses. To see how low the concentration must be to avoid alignment of the rods due to hard-core interactions, consider tobacco mosaic virus (TMV), with major axis $a_p=300\,nm$, minor axis $b_p=20\,nm$, and aspect ratio of $p=15$.  The TMV is considered to be in an isotropic phase when the volume fraction $\Phi_s \lessapprox 0.1$ \cite{oldenbourg1988orientational}, which corresponds to a concentration of approximately $C \lessapprox 0.1 \ \mathrm{g/cm}^3$.  }

% we characterise the nematic phase of a Tobacco-Mosaic-Virus (TMV) whose major axis is $a_p=300 \mathrm{nm}$, minor axis is $b_p = 20 \mathrm{nm}$ and the molecular weight is $M_w = 4.08 \times 10^{7} \mathrm{Da}$. The aspect ratio of this particle is $p=15$. 

% The volume fraction and concentration are given by, $$ \Phi_c = \frac{N \pi b_p^2 a_p}{4 V}, \qquad C = \frac{N M_w}{V N_a}$$ where $N$ is the number of particles per unit volume $V$ and $N_a$ is the Avogadro's number. The TMV is considered to be in an isotropic phase when the $\Phi_s \lessapprox 0.1$ \cite{oldenbourg1988orientational}. Therefore the relation between the phase and the concentration is $$ \Phi_s = \frac{N_a \pi a_p b_p^2}{4 M_w} C$$ and from the figure \ref{fig:tmv} we can see that the dilute limit holds true for $C \lessapprox 0.1 g/cm^3$.

% \begin{figure}[h]
%     \centering
%     \includegraphics[width=0.5\textwidth]{tmv_phase.eps}
%     \caption{Relation between the phase and concentration of TMV.}
%     \label{fig:tmv}
% \end{figure}
% }

%Monte Carlo methods are used in a range of applications, including dispersion problems ~\cite{aminian2016boundaries,peng2020upstream}.
\label{sec:MC}
\subsection{Method}

We employ a Monte Carlo method to simulate the advection, translational  diffusion, and rotational diffusion of rods in a two-dimensional channel with Poiseuille flow $\mathbf{u}(y)$ where
\begin{equation}\label{vel_vector}
    \mathbf{u}(y) = U \left[ 1 - \left(\frac{y}{a}\right)^2  \right] \hat{x} = u(y) \hat{x}.
 \end{equation} 
 We write the governing equations as stochastic differential equations since these equations directly correspond to our numerical approach (see also~\cite{aminian2016boundaries}), but our equations could equally well be written in Langevin form~\cite{han2006brownian}. The translational displacements of the particle in the laboratory frame are given by
 \begin{eqnarray}\label{MC_Dim_1}
   d x &=& u(y(t))  dt + \sqrt{2 D_{\parallel}} d W_{\parallel} \cos{\theta(t)} - \sqrt{2 D_{\perp}} d W_{\perp} \sin{\theta(t)} \\ d y &=&  \sqrt{2 D_{\parallel}} d W_{\parallel} \sin{\theta(t)} + \sqrt{2 D_{\perp}} d W_{\perp} \cos{\theta(t)}.
\end{eqnarray}
The white noise increments $d W_{\perp}$ and $d W_{\parallel}$ have zero mean,  variance $\review{dt}$, and are independent at different times.  Similarly, the stochastic differential equation for the particle orientation is
\begin{equation}\label{MC_Dim_2}
    d \theta = \omega(y(t),\theta(t)) dt + \sqrt{2 D_{\theta}} dW_\theta, 
\end{equation}
with the rotational velocity given by equation~(\ref{JO}) for the flow~(\ref{vel_vector}),
\begin{equation}\label{JO_Poiseulle}
    \omega(y,\theta) = -2 U \frac{y}{a^2} \frac{p^2 \sin^2{\theta} + \cos^2{\theta}}{p^2 + 1}.
 \end{equation}
The white noise increments $dW_\theta$ in equation \eqref{MC_Dim_2} have zero mean and variance $dt$.
\par
We non-dimensionalize equations \eqref{vel_vector}--\eqref{JO_Poiseulle} via $\mathbf{\Tilde{x}} = \mathbf{x}/a$, $\Tilde{t} = t/t_d=t/\left(a^2/\Bar{D}\right)$, $\Tilde{u} = u/U$, $\Tilde{\omega} = \omega a/U$ and $\Tilde{\mathsf{D}}= \mathsf{D}/\Bar{D}$. Dropping the tildes, equations \eqref{MC_Dim_1}--\eqref{MC_Dim_2} become 
\begin{eqnarray}
    d x &=& \mathrm{Pe} \, u(y(t)) \, dt + \sqrt{2 D_{\parallel}} d W_{\parallel} \cos{\theta(t)} - \sqrt{2 D_{\perp}} d W_{\perp} \sin{\theta(t)} \\ d y &=&  \sqrt{2 D_{\parallel}} d W_{\parallel} \sin{\theta(t)} + \sqrt{2 D_{\perp}} d W_{\perp} \cos{\theta(t)} \label{ND_MC_Rods} \\
    d \theta &=& \mathrm{Pe}\, \omega(y(t),\theta(t)) dt + \sqrt{2 \frac{\mathrm{Pe}}{\mathrm{Pe_r}}} dW_\theta,
\end{eqnarray}
where $\mathrm{Pe}=Ua/\Bar{D}$ (equation~(\ref{Pe_def_spheres})) and $\mathrm{Pe_r}=U/(a D_\theta)$ (equation~(\ref{Rotational Peclet})).
% The system is characterized by three non dimensional numbers, the Peclet number defined based on the orientationally averaged diffusivity
% \begin{equation}
%     \mathrm{Pe}=\frac{U a}{\Bar{D}},
% \end{equation}
% the rotational Peclet number defined as,
% \begin{equation}
%     \mathrm{Pe_r} = \frac{U}{a D_{\theta}},
% \end{equation}
% and the aspect ratio of the particles $p$. 
% \par
% We anticipate the rotational Peclet number role to play a key conceptual role in our problem as it directly quantifies the role of fluid shear to rotational diffusion. 
The initial condition for the simulation is $n = 10^6$ particles uniformly distributed across $y$ and across all orientations $\theta$, but with a Gaussian distribution in $x$ of unit variance centered at $x=0$. The particles are non-interacting and evolve independently. The boundary conditions at the walls are billiard-like. For a collision at a wall, the center-of-mass trajectory of a particle has an angle of incidence equal to the angle of reflection, and the orientation is assumed unchanged. \review{The influence of this orientation collision condition on the global long time statistics of the Monte Carlo simulation is examined in detail in Appendix A}.  To solve the governing equations for each particle, we  use Euler time-stepping with a dimensionless time-step of $dt=4 \times 10^{-5}$. Consequently, the typical magnitude of the white noise is therefore much less than the width of the channel, so that it is exceedingly rare for there to be more than one wall collision in a time step. Since the Monte Carlo evolution is implemented  at each time step on all the particles, the code is parallelized over many CPUs to reduce computational time.  The complete Monte Carlo simulation code is included as Supplemental Material. 
Although it is a slow method with a convergence rate that scales with $1/\sqrt{n}$, the gridless stochastic differential equation approach is convenient for combining and capturing all statistics  ~\cite{kulkarni2016modeling,lapeyre2003introduction,kloeden2012numerical}. 

We compute ensemble averages by carrying out $r$ runs of the motion of the $n$ particles. For the results reported here we take $r = 100$. The time-dependent mean and the variance of the $x$  components of all $n/r$ particles in a given run are calculated as
\begin{equation}
    \mu_i(t) = \frac{r}{n} \sum_{j=1}^{n/r} x_{i,j} (t) \qquad\text{and}\qquad
    \sigma_i^2 (t) = \frac{r}{n} \sum_{j=1}^{n/r} \left( x_{i,j}(t) - \mu_i(t) \right)^2
\end{equation}
{and then these quantities are averaged over all runs yielding} 
\begin{equation}
    \Bar{\mu}(t)= \frac{1}{r} \sum_{i=1}^{r} \mu_i(t) \qquad\text{and}\qquad
    \Bar{\sigma}^2 (t) = \frac{1}{r} \sum_{i=1}^r \left( \sigma_i^2 (t) + \left(\mu_i (t) - \Bar{\mu}(t)\right)^2\right).
\end{equation}
When $n\rightarrow\infty$, the mean particle speed and dispersion coefficient are given by
\begin{equation}\label{slopevar}
   u_m= \left. \frac{d \Bar{\mu}}{dt}\right|_{t\to\infty}
\qquad\text{and}\qquad
   \kappa= \frac{1}{2} \left. \frac{d \Bar{\sigma}^2}{dt}\right|_{t\to\infty},
\end{equation}
respectively.
% By combining statistics, computing the mean position and variance of all particles at each time step, the plot of the variance with time can give us the effective diffusivity.
% \par
% All the essential data from each CPU instance was stored, like mean, variance, time vector, position, and orientation of the last time step, which made combining statistics of different runs easy.  There are $N/n=r$ independent runs completed for a given set of parameters.  For the $i$th computational run containing $n$ particles, the mean of the run defined as
% \begin{equation}
%     \mu_i(t) = \frac{1}{n} \sum_{j=1}^{n} x_{i,j} (t)
% \end{equation} and variance as
% \begin{equation}
%     \sigma_i^2 (t) = \frac{1}{n} \sum_{j=1}^{n} \left( x_{i,j}(t) - \mu_i(t) \right)^2,
% \end{equation}
% each of which are stored at each time step, for each run.  Due to the statistical independence of each run, the mean and the variance for the all the different $r$ runs can be combined as,
% \begin{equation}
%     \Bar{\mu}(t) = \frac{1}{r} \sum_{i=1}^{r} \mu_i(t)
% \end{equation}
% and,
% \begin{equation}
%     \Bar{\sigma}^2 (t) = \frac{1}{r} \sum_{i=1}^r \left( \sigma_i^2 (t) + \left(\mu_i (t) - \Bar{\mu}(t)\right)^2\right).
% \end{equation}
% The effective diffusivity of the particles is mathematically defined as, 
% \begin{equation}\label{slopevar}
%     \left. \frac{d \Bar{\sigma}^2}{dt}\right|_{t\to\infty} = \kappa
% \end{equation}
% For the case of spherical particles in a 2-D channel flow equation \ref{kappa_s} is the effective diffusivity equation. 
In practice, there are transients in the dispersion that decay after a dimensionless time of approximately $0.25 t_d$~\cite{dutta2006effect,aminian2016boundaries}. 
Therefore, to calculate the effective diffusivity, we fit the computed variance to an expression of the form
\begin{equation}
%\Bar{\sigma}^2(t) = s - a_2 + a_1 t + a_2 \exp(-a_3t),
{\sigma}^2(t) = s - a_1 (1 - e^{-a_2t}) + 2 \kappa t,
\end{equation}
using a least-squares method, where $s=1$ is the initial variance in $x$.
%where $s$ is the initial variance as defined by the initial conditions, $a_2$ and $a_3$ are fitting constants for the transient growth phase %of the variance. $a_1$ is the slope of the linear phase. Substituting the fitting expression into equation \ref{slopevar},
% \begin{equation}
%     \kappa = a_1.
% \end{equation}
%We obtain the mean speed of the particles in a similar manner. We fit a curve of the form, 
Likewise, we fit the mean speed of the particles to
\begin{equation}
    {\mu} (t) = b_0 + b_1 e^{-b_2 t} + u_m t,
\end{equation}
to find the mean speed $u_m$ at long times.
% where $b_0$ is the statistical offset and $b_2$ and $b_3$ are fitting constants corresponding to the transient growth phase and $b_1$ is the slope of the linear phase. The mean speed of the particles at long times ($t\to\infty$) defined $\Bar{U}_p$ is thus taken as 
% \begin{equation}
%     \Bar{U}_p = b_1.
% \end{equation}

%%%Results
\subsection{Results}

%%%start figure
\begin{figure}[ht] %should I make the grey for Pe_r=10 lighter? Will take one second.
    \centering
    \includegraphics[width=1\textwidth]{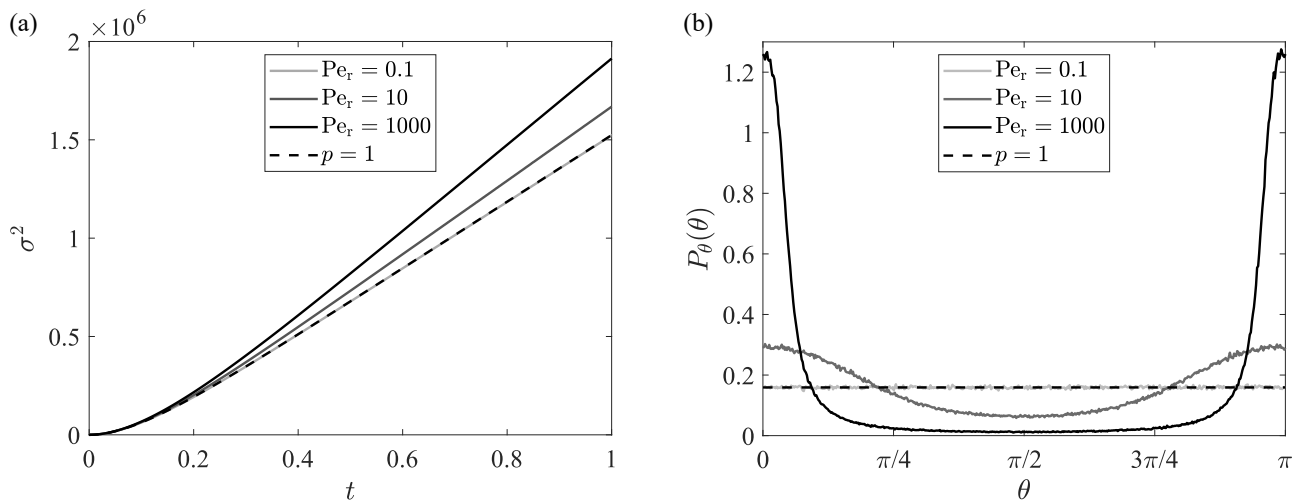} 
    \caption{(a) Monte Carlo results for the variance of the $x$-position of  ellipsoidal particles ($p=1000$) and spherical particles at $\mathrm{Pe}=10^4$ for different $\mathrm{Pe_r}$ as a function of dimensionless time. The rods disperse along $x$ like spheres when rotational Brownian motion dominates ($\mathrm{Pe_r}\ll1$). The dispersion of rods is larger when shear dominates ($\mathrm{Pe_r} \gg 1$), i.e., when the rod's orientations follow Jeffery orbits. The complete theoretical prediction for the variance of spherical particles in a two-dimensional channel has been reported previously and is also shown here for comparison (dashed line)~\cite{bernardi2018space}. (b) Monte Carlo results for the orientational distribution $P_{\theta} (\theta)$ for particles over the channel's length. The rods spend more time aligned with the flow direction when $\mathrm{Pe_r} \gg 1$.}
    \label{fig:Compare rods and spheres variance}
\end{figure}
%%%end figure

In Taylor's original picture, flow enhances spreading due to differences in the flow speed across the channel.  Our simulations reveal that this enhancement is, in fact, {\it increased} for rod-like particles, as shown in Figure~\ref{fig:Compare rods and spheres variance}(a). Physically, spherical particles rotate uniformly in shear. However, rod-like particles have a non-uniform rotation rate (Figure~\ref{fig:curvesJO_D}(b)), and thus spend more time aligned with the flow than perpendicular to the flow. This alignment effect becomes stronger as the rotational Peclet number, $\mathrm{Pe_r}$, increases (see Figure~\ref{fig:Compare rods and spheres variance}(b)).
\begin{figure}[ht]
    \centering
    \includegraphics[width=1\textwidth]{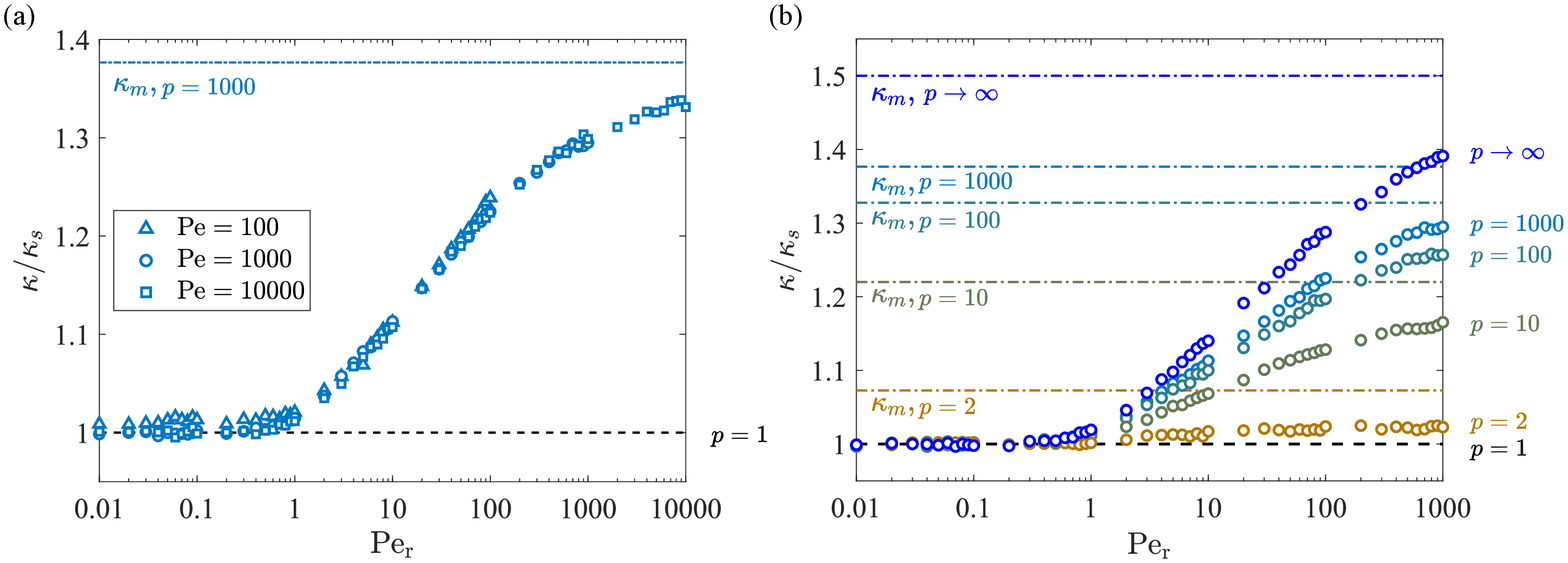}
    \caption{(a) Effective diffusivity, $\kappa$, of rod shaped particles, normalized by the effective diffusivity for spheres, as a function of $\mathrm{Pe_r}$ for various values of $\mathrm{Pe}$. The triangles represent Monte Carlo simulations for $\mathrm{Pe}=100$, the circles represent $\mathrm{Pe}=1000$, and the squares represent $\mathrm{Pe}=10000$. (b) Variation of normalized effective diffusivity with aspect ratio $p$. In both panels, the dash-dotted line represents the maximum theoretical value of dispersion for the corresponding aspect ratio. The maximum possible dispersion constant is estimated when all rod shaped particles are aligned in the direction of the flow and is defined as per equation \eqref{kappa_max_disp}.}
    \label{fig:Diff Pe_r}
\end{figure} 
%%% figure end
For small values of $\mathrm{Pe_r}$, the rod shaped particles rotate randomly and spread identically to spherical particles. As the shear rate increases, the strong alignment in the direction of the flow causes the perpendicular ``side" of the particles (which has a lower diffusivity than spherical particles) to diffuse across the shear layers. %Furthermore, for rods with a high aspect ratio $p \ge 10$, the alignment is strong, causing an enhanced effect in the dispersion. 
We can be somewhat more quantitative by noting that the effective lateral diffusivity $D_y$ (defined more precisely in the next section) is smaller for rods than spheres. Thus, since we expect $\kappa^\prime\propto U^2a^2/D_y$, and since $\kappa^\prime_s\propto U^2 a^2/D$, we have
% The alignment causes the perpendicular side of the particles, (which has a lower diffusivity compared to spherical particles) to diffuse across the shear layers. This effective lowered value of average lateral molecular diffusion constant enhances the longitudinal dispersion. Mathematically, the effective dispersion constant is inversely proportional to longitudinal diffusion constant \cite{taylor1953dispersion}. Dimensionally, we define  
% \begin{equation}
%     \kappa' \sim \frac{U^2 a^2}{D_y} %need new term or use $D_y$ %consider dimensional just one d_y. Should I explain it in the form?
% \end{equation}
% where prime denotes the dimensional form of the dispersion factor. In non dimensional form where the $\kappa'$ and $D_y$ are non dimensionalised using $\Bar{D}$ such that $\Tilde{D_y} = D_y/\Bar{D}$, we obtain,
% \begin{equation}
%     \kappa \sim \frac{U^2 a^2}{\Bar{D}^2 D_y} =\frac{\mathrm{Pe}^2}{D_y}
% \end{equation}
% where the tildes have been dropped in the non dimensional forms. Normalising by the dispersion factor with the dispersion for spherical particles the above equation becomes,
\begin{equation}\label{scaling_D_y}
    \frac{\kappa}{\kappa_s}=\frac{\kappa^\prime}{\kappa_s^\prime} \sim \frac{D}{D_y}.
\end{equation}

Our Monte Carlo results for the effective diffusivity are shown as a function of rotational Peclet number in Figure~\ref{fig:Diff Pe_r} for various values of the Peclet number (Figure \ref{fig:Diff Pe_r}(a)) and aspect ratio (Figure \ref{fig:Diff Pe_r}(b)). {\color{black} Since all of the curves collapse in} Figure \ref{fig:Diff Pe_r}(a), we can conclude that the $\mathrm{Pe}^2$ scaling holds for rod-shaped particles at $\mathrm{Pe} \gtrsim 100$, as is the case for spherical particles [Eq.(\ref{kappa_s})]. Figure \ref{fig:Diff Pe_r}(b) demonstrates that at low $\mathrm{Pe_r}$, rod shaped particles behave like spherical particles as rotational diffusion dominates, and the rods are oriented randomly. Furthermore, as the $\mathrm{Pe_r}$ increases, we see the rods tend to align themselves in the direction of the flow due to their Jeffery's orbit and ultimately spread more. Rods with larger aspect ratios have a stronger alignment and a lower perpendicular diffusion constant ($D_{\perp}$), and thus spread more. 

%Enhanced spreading of rod-shaped particles as compared to spherical can also be explained qualitatively with the following argument. Particles along the edges of the channel have a lower velocity and lower longitudinal diffusion constant compared to spherical particles, making it harder for them to diffuse into the center of the channel and \emph{vice versa}. Simultaneously, the particles in the center are advected further downstream, and hence the width of the solute plug becomes larger for rods compared to spherical particles 

\begin{figure}[ht]
    \centering
    \includegraphics[scale=0.65]{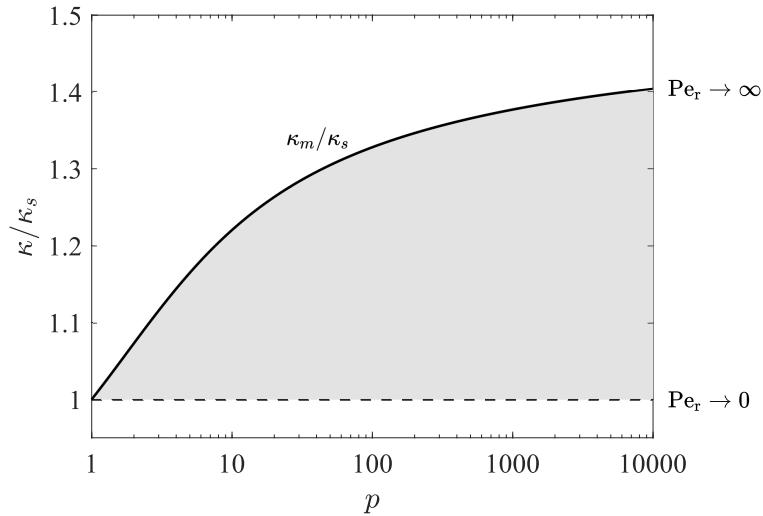}
    \caption{The maximum possible dispersion $\kappa_m$ normalized by $\kappa_s$ as a function of aspect ratio $p$. The shaded area corresponds to the region of possible values of $\kappa/\kappa_s$ for all $p$ and $\mathrm{Pe_r}$.}
    \label{fig:k_max}
\end{figure}

For a given set of parameters, the maximum possible value of dispersion anticipated, $\kappa_m$, can be estimated by simply assuming all of the particles maintain perfect alignment with the flow. Thus $D_{y} = D_{\perp}$ and
\begin{equation}\label{kappa_max_disp}
    \frac{\kappa_{m}}{\kappa_s} = \frac{\bar{D}}{D_{\perp}}.  
\end{equation}
The ratio $\kappa_{m}/\kappa_s$ depends solely on the aspect ratio of the rod, $p$, and increases monotonically from $\kappa_{m}/\kappa_s=1$ when $p=1$ (spherical particle) to $\kappa_{m}/\kappa_s=3/2$ as $p\rightarrow \infty$ (slender body limit). Figure \ref{fig:k_max} shows the maximum possible dispersion as a function of the aspect ratio and allows us to define a region where we expect to find values of $\kappa$ in practice.

We note that the results of the Monte Carlo simulations presented here only make physical sense for $\mathrm{Pe_r} < \mathrm{Pe}$, as we now describe.
The ratio of \review{$\mathrm{Pe_r} = U/a D_\theta$ and $\mathrm{Pe} = U a/\bar{D}$} is the ratio of the rotational and translational diffusive time scales
\begin{equation}\label{particle channel}
    \frac{\mathrm{Pe_r}}{\mathrm{Pe}} = \frac{\Bar{D}}{a^2 D_\theta} \review{\sim \frac{a_p^2}{a^2}    \ll 1.}
\end{equation}
% \review{
% From equation \ref{eqDpara} - \ref{D_thetaeq},
% \begin{equation}
% \label{eqn:ratio_time_scales}
%     \frac{\mathrm{Pe_r}}{\mathrm{Pe}} \sim \frac{a_p^2}{a^2}    \ll 1.
% \end{equation}
%\frac{\left(p^2+1\right) \left(\sqrt{p^2-1} p-\left(2 p^2-3\right) \log \left(\sqrt{p^2-1}+p\right)-\left(2 p^2-1\right) \log \left(\frac{\sqrt{p^2-1}+p}{p-\sqrt{p^2-1}}\right)\right)}{6 \left(\sqrt{p^2-1} p+\left(1-2 p^2\right) \log \left(\sqrt{p^2-1}+p\right)\right)}
Since we focus on the physically relevant regime where $a_p \ll a$, this condition suggests restricting our attention to $\mathrm{Pe_r} \ll \mathrm{Pe}$, a fact we will exploit in the following section to derive semi-analytical expressions for the dispersion coefficient, $\kappa$, and mean particle speed, $u_m$. \review{For example, an elongated TMV particle with $a_p=300 \ \text{nm}$ and $p=15$ in a channel with $a=2 \ \mu \text{m}$ flowing in water with a velocity $U=1 $ mm/s, will have $\mathrm{Pe_r} = 10$ and $\mathrm{Pe} = 750$, and is therefore likely to exhibit enhanced dispersion.}

%\color{red} TO AJAY: Can you use the parameters corresponding to the TMV, since we already used that as a specific example? Done, Ajay: I Can also add an example when it is safe to assume the nano particle to be spherical.}   

%-------------THEORETICAL ANALYSIS-----------------------%

\section{Theoretical Analysis} 
\label{sec:simp_theory}
In this section, we generalize Taylor's continuum analysis of the dispersion of spherical particles in a shear flow to ellipsoidal particles. We write the Fokker-Planck equation for the probability density function for the particles' positions and orientations. We then use an asymptotic analysis to determine an effective one-dimensional transport equation with an effective dispersion coefficient and the longitudinal transport speed analogous to equation \eqref{eqn:Taylor_result}.  \review{We note that alternative analytical approaches could be employed to arrive at similar quantities of interest \cite{frankel1989foundations,jiang2019dispersion}. In the present work, we restrict our attention to the physically relevant regime where $\mathrm{Pe_r} \ll \mathrm{Pe}$ which facilitates a simpler analysis in the spirit of Taylor's original calculation, while still demonstrating excellent quantitative agreement with the full Monte Carlo simulation.}   

\subsection{Conservation equation: the Fokker-Planck model}
\label{sec:FP}
We define the probability distribution by $P(\mathbf{x},\theta,t) = \mathcal{C}(\mathbf{x},\theta,t)/N$, where $\mathcal{C}(\mathbf{x},\theta,t)\Delta x\Delta y\Delta\theta$ gives the number of solute particles in a small region of dimensions $\Delta x\Delta y\Delta\theta$ about $(x,y,\theta)$ at time $t$, and $N$ is the total number of particles. Conservation of particles implies the probability distribution obeys the Fokker-Planck equation
\begin{equation}\label{flux}
    \frac{\partial P}{\partial t} + \boldsymbol{\nabla}\cdot \mathbf{J} + \frac{\partial}{\partial \theta}  {J}_\theta  = 0,
\end{equation}
where the translational flux is $\mathbf{J}$ and the rotational flux is $J_\theta$. Each of these fluxes has contributions from both advection and diffusion:
\begin{equation}
    \mathbf{J}=\mathbf{u} P \review{-} \mathsf{{D}}\cdot\boldsymbol{\nabla} P, \qquad\text{and}\qquad
    J_\theta=\omega P \review{-} D_\theta\frac{\partial P}{\partial \theta},
\end{equation}
with $\mathbf{u}$ given by the flow in equation (\ref{vel_vector}), and $\omega$ given by the rotation rate of the Jeffery orbit in equation (\ref{JO}).  The diffusion tensor $\mathsf{D}$ is given by~\cite{brenner1974transport}
\begin{equation}
    \mathsf{{D}}(\theta) = \mathbf{e} \ \mathbf{e} D_{\parallel} + (\mathsf{{I}} - \mathbf{e} \ \mathbf{e})D_{\perp},
\end{equation}
where $\mathbf{e}=\cos \theta \ e_x + \sin \theta \ e_y.$ 
% The conservation equation has a translational and rotational flux. The translational diffusion depends on the orientation of the particles \cite{brenner1974transport}. Since the rotational and translational components can be decoupled for our system, the diffusion tensor is of the form 
% \begin{equation}
%     \mathsf{{D}}(\theta) = \mathbf{e} \ \mathbf{e} D_{\parallel} + (\mathsf{{I}} - \mathbf{e} \ \mathbf{e})D_{\perp},
% \end{equation}
% where $\mathbf{e}\tp{=(\cos\theta,\sin\theta).}$ 
%is 
%the unit vector aligned along the axis of symmetry of the semi major axis. $\underline{e} \ \underline{e}$ is the dyadic product and $\underline{\underline{I}}$ is the identity matrix. Therefore for our choice of coordinate system,
% %\begin{equation}\label{unit_vector}
%     \underline{e} = \begin{bmatrix} \cos{\theta} & \sin{\theta} \end{bmatrix}
% \end{equation}
% and,
% \begin{equation}\label{dydadic_product}
%     \underline{e} \ \underline{e} = \begin{bmatrix} \cos^D{\theta} & \sin{\theta} \cos{\theta} \\ \sin{\theta} \cos{\theta} & \sin^2{\theta} \end{bmatrix}.
% \end{equation}
In the $xy$ (laboratory) basis, the components of the translational diffusion tensor are
\begin{equation}\label{difftensor2}
        \begin{bmatrix} D_{xx}(\theta)  & D_{xy}(\theta)  \\ D_{xy}(\theta)  & D_{yy}(\theta) \end{bmatrix} = \begin{bmatrix} D_{\parallel}\cos^2{\theta} + D_{\perp} \sin^2{\theta} & (D_{\parallel} - D_{\perp}) \sin{\theta} \cos{\theta}  \\ (D_{\parallel} - D_{\perp}) \sin{\theta} \cos{\theta}  & D_{\parallel}\sin^2{\theta} + D_{\perp} \cos^2{\theta} \end{bmatrix}.
\end{equation} 
%\review{TP: eqn (30) I would flip the order of the matrices, and not set tensor (D)  equal to the components, AHK: Didn't understand. !Done!}
% and the orientationally averaged diffusivity is,
% \begin{equation}
%     \Bar{D} = \frac{tr(\underline{\underline{D}})}{2} = \frac{D_{\parallel} + D_{\perp}}{2}.
% \end{equation}
% The conservation equation for a probability density function $P(x,y,\theta,t)$ is 
% \begin{equation}\label{flux}
%     \frac{\partial P}{\partial t} + \boldsymbol{\nabla}_{x,y} \cdot \mathbf{J}_{x,y} + \frac{\partial}{\partial \theta} \left( {J}_\theta \right) = 0,
% \end{equation}
% where $\mathbf{J}_{x,y}$ is the flux in the translational space consisting of the advective and the diffusive flux, $ {J}_{\theta}$ is the flux in the orientational space consisting of an advective flux or the Jeffery's Orbit and the rotational diffusion flux. 
%The above equation is a Fokker-Planck equation and can be written as
Thus, the conservation equation~(\ref{flux}) can be written as
\begin{equation}\label{eqn:masterdim}
\begin{aligned}
    \frac{\partial P}{\partial t} = -u(y) \frac{\partial P}{\partial x} + D_{xx}(\theta) \frac{\partial^2 P}{\partial x^2} + 2 D_{xy} (\theta) \frac{\partial^2 P}{\partial x \partial y} + D_{yy}(\theta) \frac{\partial^2 P}{\partial y^2} + D_{\theta} \frac{\partial^2 P}{\partial \theta^2} - \frac{\partial}{\partial \theta}\left[ \omega(y,\theta) P \right].
\end{aligned}
\end{equation}
\review{The symmetry of the rod-shaped particles makes the probability distribution periodic in $\theta$, with $P(\mathbf{x},\theta + \pi,t) = P(\mathbf{x},\theta,t)$. We also demand no-flux boundary condition at the walls \review{~\cite{ezhilan_saintillan_2015,nitsche1997shear}, hence
\begin{equation}
    \label{eqn:no_flux_BC_1}
     \left(\mathbf{J}\cdot\hat{y}\right) = D_{xy} (\theta) \frac{\partial P}{\partial x }+ D_{yy} (\theta) \frac{\partial P}{\partial y} = 0\qquad\text{at}\qquad y = \pm a.
 \end{equation}}}

%where $D_{\theta}$ is the rotational diffusion constant. $D_{ij}$ are the elements of the diffusion tensor obtained as per equation \ref{difftensor2}. The velocity profile and the Jeffery's orbit are defined as per equation \ref{vel_vector} and \ref{JO_Poiseulle} respectively. 

% There are three imposed conditions: periodicity in the orientational space
% \begin{equation}
%     P(x,y,\theta=0,t)=P(x,y,\theta=\pi,t),
% \end{equation}  
% a no-flux boundary condition at the walls
% \begin{equation}\label{no_flux_BC_1}
%     \left(\mathbf{J}\cdot\hat{y}\right)|_{y=\pm a} = D_{yy} (\theta) \left. \frac{\partial P}{\partial y} \right|_{y=\pm a} = 0,
% \end{equation} 
% and an initial condition 
% \begin{equation}
%     P(x,y,\theta,t=0) = \frac{1}{4 \pi} \delta(x).
% \end{equation}
% Since the particles have a periodicity of $\pi$, the angular domain can be reduced to $0$ to $\pi$ from $0$ to $2\pi$.
% \par

%We seek a transport equation, analogous to equation \eqref{eqn:Taylor_result}, for the particle concentration valid long after transverse diffusion has spread the solute across the width of the channel, specifically for time $t\gg a^2/\bar{D}$. In this regime, the solute has dispersed over a length $L\sim U t$ along the channel. Thus $L\gg U a^2/\bar{D}$, or $L/a\gg \mathrm{Pe}$. To enforce this limit, we choose $L = a/\varepsilon$, with the understanding that $\varepsilon = \mathrm{Pe_r}/\mathrm{Pe} \ll 1$, consistent with equation \eqref{eqn:ratio_time_scales}. Employing the scalings

\review{We consider dispersion of solute relative to a frame traveling with an \emph{a priori} unknown mean particle speed $u_m$, prompting the change of variables $X = x - u_m t$. In classical Taylor dispersion for spherical particles, $u_m$ coincides with the mean speed of the flow, specifically $u_m = 2U/3 $. In preparation for the asymptotic procedure outlined in \S\ref{sec:dispCoeff_meanPS}, equations \eqref{eqn:masterdim} and \eqref{eqn:no_flux_BC_1} in the Lagrangian frame are non-dimensionalized \emph{via} the following scalings (as in the Monte Carlo):
$$
t = \frac{a^2}{\bar{D}}\hat{t},\qquad u = U\hat{u},\quad D_{ij} = \bar{D}\hat{D}_{ij},\qquad (X, y) = a(\hat{X},\hat{y}),\qquad \omega = \frac{U}{a}\hat{\omega}.
$$
Employing these scalings leads to the dimensionless conservation equation
\begin{subequations}
\label{eqn:FP_nondims}
 \begin{equation}
     \label{eqn:FP_nondim1}
     \varepsilon \frac{\partial P}{\partial \hat{t}} = -\mathrm{Pe_r}  (\hat{u}(\hat{y}) - \hat{u}_m)\frac{\partial P}{\partial \hat{X}} + \varepsilon \hat{D}_{xx}(\theta)\frac{\partial^2 P}{\partial \hat{X}^2} + 2\varepsilon \hat{D}_{xy}(\theta)\frac{\partial ^2 P}{\partial \hat{X}\partial \hat{y}} + \varepsilon \hat{D}_{yy}(\theta)\frac{\partial^2 P}{\partial \hat{y}^2} + \frac{\partial^2 P}{\partial\theta^2} - \mathrm{Pe_r} \frac{\partial}{\partial\theta}\left[\omega(\hat{y},\theta) P\right]
     \end{equation}
     and zero-flux boundary condition
     \begin{equation}
     \label{eqn:FP_nondim2}
     \varepsilon \hat{D}_{xy} (\theta)\frac{\partial P}{\partial \hat{X} } + \hat{D}_{yy} (\theta) \frac{\partial P}{\partial \hat{y}} = 0\qquad\text{at}\qquad \hat{y} = \pm 1,
 \end{equation}
\end{subequations}
where we have defined $\varepsilon = \mathrm{Pe_r}/\mathrm{Pe} \ll 1$, consistent with the physically relevant regime [Eq. \eqref{particle channel}]. Recall that our focus in the present work is on Taylor's regime wherein $\mathrm{Pe}\gg 1$.  Henceforth, we drop the hat decorations denoting dimensionless quantities to reduce clutter.

\subsection{The dispersion coefficient and mean particle speed}
\label{sec:dispCoeff_meanPS}

Our goal is to derive an effective transport equation for long times, analogous to equation \eqref{eqn:Taylor_result}, for the particle concentration valid long after transverse diffusion has spread the solute across the width of the channel. Taylor's original result \cite{taylor1953dispersion} similarly describes the concentration evolution in long time, specifically after the dispersing plug's length is much larger than $U t_d = a \, \mathrm{Pe}$.  Consistent with Taylor's condition and our assumptions hitherto, we introduce the slow space variable $\xi = \varepsilon^2 X$ for our modified Taylor dispersion analysis.  Our Monte Carlo simulations indicate an enhanced dispersion factor that scales with $\mathrm{Pe}^2$ (as in classical Taylor dispersion) and when combined with the selected slow space variable scaling, suggest a long time scale $T = \varepsilon^2 t$.  Finally, we observe that \eqref{eqn:FP_nondim1} suggests that the timescales for the different relaxation processes are well-separated when $\varepsilon\ll 1$ and $\mathrm{Pe_r} = \mathcal{O}(1)$, with the orientational dynamics occurring most rapidly.   
In the long-time regime considered here, we assume that these rotational degrees of freedom have relaxed to their steady-state values \cite{nitsche1997shear}. Amalgamating these considerations suggests that we seek solutions of the form
%Diffusion across the channel and advection along the channel have characteristic dimensionless rates of order unity. The dimensionless rate for diffusion across the channel given a gradient along the channel, or vice versa, is $\sim \varepsilon$. Finally, the diffusion along the channel given a gradient along the channel is the slowest process, with dimensionless rate $\sim\varepsilon^2$. We are thus prompted to define the long, diffusive time scales $t_1 = \varepsilon t$ and $t_2 = \varepsilon^2 t$. Further, we note that we can consider every $x$-derivative in \eqref{eqn:FP_nondims} to come with a factor of $\varepsilon$, indicating the small gradient of solute concentration along the channel. Thus, it is natural to define the slow spatial variable $X=\varepsilon x$. Altogether, these considerations suggest that we seek solutions of the form
\begin{equation}
\label{eqn:P_ansatz}
    P(x,y,\theta,t) = \frac{1}{N} g(\theta; y)\mathcal{C}(\xi,y,T),
\end{equation}
where $g$ represents the orientational distribution of the particles at each shear layer, $y$, and $\mathcal{C}$ is the net concentration of particles at position $(\xi,y)$. 
%\sjt{[[We have a few variables that we call the ``concentration'' - we ought to be careful to distinguish between them.]]} 
We then expand the concentration, $\mathcal{C}$, and unknown mean particle speed, $u_m$, in powers of $\varepsilon$ as follows
\begin{equation}
\label{eqn:C_expansion}
    \mathcal{C}(\xi,y,T) = \mathcal{C}^{(0)}(\xi,y,T) + \varepsilon \mathcal{C}^{(1)}(\xi,y,T) + \varepsilon^2 \mathcal{C}^{(2)}(\xi,y,T) + \mathcal{O}(\varepsilon^3),\qquad u_m = u^{(0)}_m + \varepsilon u^{(1)}_m + \mathcal{O}(\varepsilon^2).
\end{equation}

After inserting the expansions \eqref{eqn:C_expansion} into equations \eqref{eqn:FP_nondims} and gathering like powers of $\varepsilon$, at leading order we find the following periodic boundary-value problem for $g$:

\begin{equation}
     \label{eqn:g_bvp}
          \frac{\partial^2 g}{\partial\theta^2} - \mathrm{Pe_r}\frac{\partial}{\partial\theta}\left(\omega(y,\theta) g\right) = 0,\qquad \int_0^{2\pi} g \ \text{d}\theta = \langle g\rangle = 1,
 \end{equation}
which is solved using a truncated Fourier series of the form \cite{nitsche1997shear}
\begin{equation}
\label{eqn:g_Fourier}
g = \frac{1}{2\pi} + \sum_{n = 1}^{M} \left\{a_n(y)\cos(2n\theta) + b_n(y)\sin(2 n \theta)\right\}.
\end{equation}
To solve for the Fourier coefficients $a_n(y)$ and $b_n(y)$, we insert the Fourier series \eqref{eqn:g_Fourier} into equation \eqref{eqn:g_bvp}, imposing the differential equation at every point $\theta_i = \pi i/I$ where $i = 1,\ldots,I.$ The result is an overdetermined, linear system of dimension $I\times 2M$. For each value of $y_k = -1 + 2k/K$, where $k = 0,\ldots,K$, the solution vector containing the Fourier coefficients was found by a standard QR least-squares algorithm in MATLAB \cite{trefethen1997numerical}. For the computations reported here, we take $I = 501$, $M = 100$, and $K = 1001$, providing more-than-sufficient accuracy for all values of $\mathrm{Pe_r}$ reported here. 

%\review{[[We don't really discuss Figure 7(a) here. I suggest either saying more or removing.]]} 
In Figure \ref{fig:g_plots}, we plot both the orientational distribution, $g$, for varying $y$ and the laterally averaged orientational distribution 
\begin{equation}
\label{eqn:g_latave_def}
    \bar{g} = \frac{1}{2}\int_{-1}^{1} g\ \text{d}y
\end{equation}
for several values of the rotational Peclet number, $\mathrm{Pe_r}$. We observe that as we move from a rotational Brownian motion to shear-dominated regime (increasing $\text{Pe}_r$), the particles have a propensity to align themselves with the flow direction, a feature quantitatively consistent with the results of our Monte Carlo simulations shown in Figure \ref{fig:Compare rods and spheres variance}. Indeed, as $\mathrm{Pe_r}\rightarrow\infty$, the solution to [Eq. \eqref{eqn:g_bvp}] develops a boundary layer near $\theta = 0$, although this limit technically violates the assumptions under which the present asymptotic analysis is valid.
\begin{figure}
        \begin{center}
        \includegraphics[width=1\textwidth]{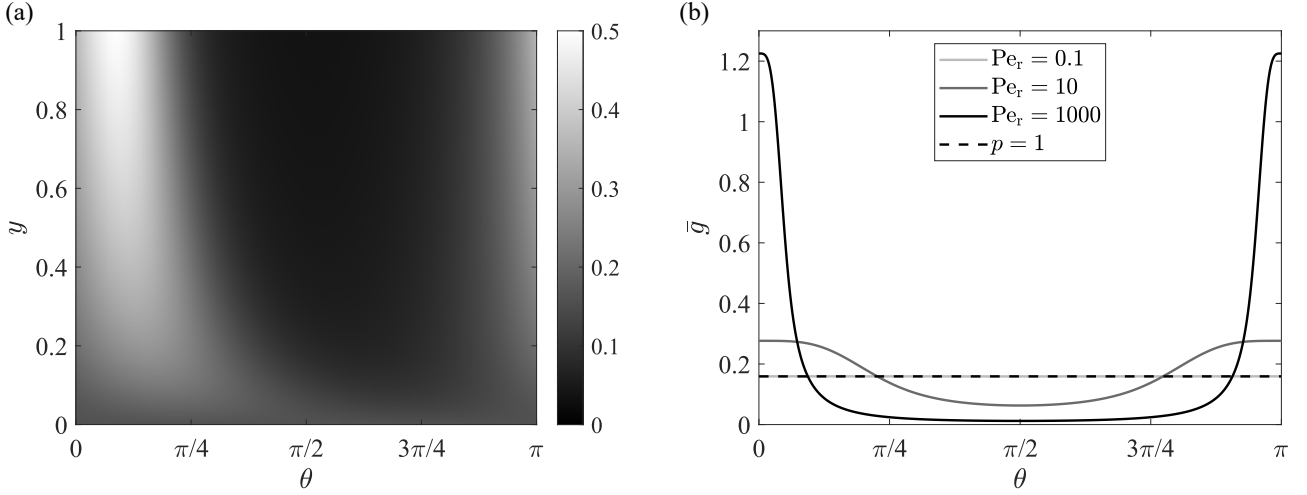}
        \caption{(a) Plot of the orientational distribution, $g(\theta;y)$, for $\mathrm{Pe_r} = 10$ and (b) plot of $\bar{g}$ versus the orientation angle, $\theta$, for several values of $\mathrm{Pe_r}$. When $\mathrm{Pe_r}$ is small and rotational Brownian motion dominates, the orientational distribution of the particles is approximately uniform; the particles have a greater propensity to align themselves with flow as $\mathrm{Pe_r}$ increases. The laterally averaged orientationally distribution compares well with the particles distribution from Monte Carlo simulations as seen in Figure \ref{fig:Compare rods and spheres variance}(b).  In both (a) and (b), we choose $p = 1000$, while the form of the rotation rate, $\omega$, allows us to restrict our plotting domain to $0\le\theta\le\pi$.}
        \label{fig:g_plots} 
        \end{center}
\end{figure}

Proceeding to $\mathcal{O}(\varepsilon)$, equation \eqref{eqn:FP_nondim1} yields
\begin{equation}
\label{eqn:C0_unaveraged}
   \frac{\partial}{\partial y} \left( D_{yy} g \frac{\partial \mathcal{C}^{(0)}}{\partial y} + D_{yy} \frac{\partial g}{\partial y} \mathcal{C}^{(0)} \right) = 0.
\end{equation}
After averaging equation \eqref{eqn:C0_unaveraged} over particle orientations, we find the following steady advection-diffusion equation
\begin{equation}
    \label{eqn:C0_averaged}
     \frac{\partial}{\partial y}\left(D_y(y) \frac{\partial \mathcal{C}^{(0)}}{\partial y} + v_d(y) \mathcal{C}^{(0)}\right)= 0,
\end{equation}
where the flux term on the left-hand side of \eqref{eqn:C0_averaged} consists of an orientationally averaged lateral diffusion coefficient and migration velocity
\begin{equation}
    \label{eqn:diff_and_vd}
D_y(y) = \langle D_{yy} g\rangle\qquad\text{and}\qquad v_d(y) = \left< D_{yy} \frac{\partial g}{\partial y} \right> = \frac{\partial D_y}{\partial y},
\end{equation}
respectively \cite{nitsche1997shear}. Hence, the solution of the advection-diffusion equation \eqref{eqn:C0_averaged} is of the form
\begin{equation}\label{eqn:solutionC0}
    \mathcal{C}^{(0)}(\xi,y,T) = \mathcal{C}_m(\xi, T)/D_y.
\end{equation}
The angle bracket notation in \eqref{eqn:diff_and_vd} is the same as that used in equation \eqref{eqn:g_bvp} to denote the orientational average of the contained quantity. Due to the form of $D_{yy}(\theta)$ given by equation \eqref{difftensor2},  $D_y$ can be expressed as 
\review{\begin{equation}
    D_y = 1 + \pi a_1 (y) \zeta \qquad\text{where}\qquad \zeta = \frac{D_{\perp} - D_{\parallel}}{D_{\perp} + D_{\parallel}}.
\end{equation}}
%AHK: Changed beta to zeta as we use beta in appendix. \\}

\review{Figure \ref{fig:drift_diffusion}(a) shows how the preferential alignment in regions of high shear near the wall reduces the lateral diffusion coefficient, in contrast to the center of the channel where $D_y = 1$ as for spherical particles.}
As shown in Figure \ref{fig:drift_diffusion}(b), particles near the center of the channel $(y = 0)$ migrate towards regions of high shear ($y = \pm 1$) with a migration velocity $v_d$. Simultaneously, the particles close to channel walls diffuse less strongly back into the bulk as shown in Figure \ref{fig:drift_diffusion}(a). 
%\review{SJT: [[I think we should reverse the order of the Figures, or alter the text, so that we don't talk about 9(b) before 9(a)]]}
\begin{figure}[ht]
     \centering
     \includegraphics[width=1\textwidth]{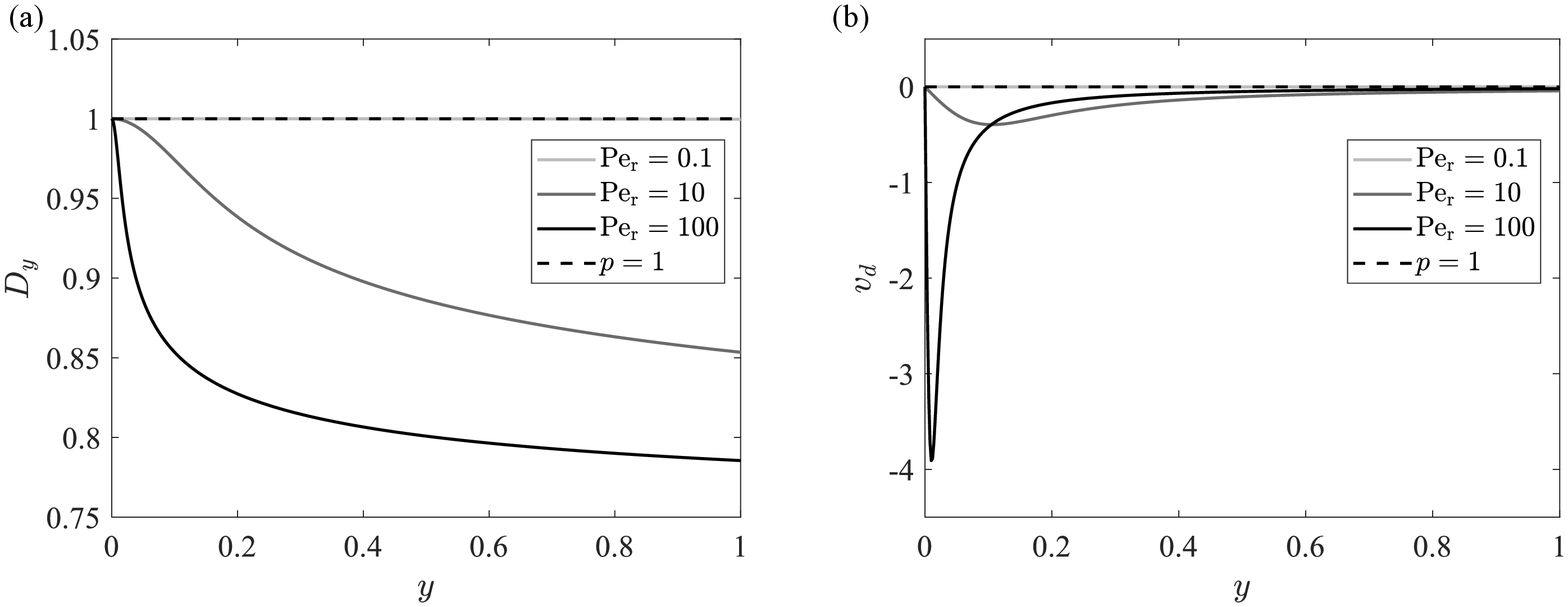}
     \caption{Plots of (a) the orientationally averaged lateral diffusion coefficient, $D_y$, and (b) the lateral migration velocity, $v_d$, for $p=1000$ as a function of the position along the width of the channel. As we move from a rotational Brownian motion (Pe$_r\ll 1$) to a shear dominated regime (Pe$_r\gg 1$), the particles migrate more strongly from $y = 0$ to the channel walls, where they simultaneously experience lower diffusion back into the bulk.}
     \label{fig:drift_diffusion}
 \end{figure}
 
After averaging over particle orientations once more and using equation \eqref{eqn:solutionC0}, at $\mathcal{O}(\varepsilon^2)$ we find from equation \eqref{eqn:FP_nondim1} 
\begin{subequations}
\begin{equation}
\label{eqn:C1_averaged}
    -D^{-1}_y \text{Pe}_r (u - u^{(0)}_m) \frac{\partial \mathcal{C}_m}{\partial \xi} + \frac{\partial^2}{\partial y^2}\left(D_y \mathcal{C}^{(1)}\right) = 0
\end{equation}
and, from equation \eqref{eqn:FP_nondim2}, the corresponding boundary condition
\begin{equation}
\label{eqn:C1_avebc}
\frac{\partial}{\partial y}\left(D_y \mathcal{C}^{(1)}\right) = 0\qquad\text{at}\qquad y = \pm 1.
\end{equation}
\end{subequations}
We obtain an expression for the leading-order mean particle speed, $u^{(0)}_m$, by first taking the lateral average of equation \eqref{eqn:C1_averaged} and then by demanding that the advective flux vanishes in the traveling frame, $\xi$. Hence, we find that
\begin{equation}
\label{eqn:um0}
u^{(0)}_m = \frac{\overline{D^{-1}_y u(y)}}{\overline{D^{-1}_y}},
\end{equation}
where the bar notation denotes the lateral average, as was introduced in equation \eqref{eqn:g_latave_def}. Finally, integrating \eqref{eqn:C1_averaged} subject to the boundary condition \eqref{eqn:C1_avebc}, we find
%\begin{equation}
%\label{eqn:CpC1}
%\mathrm{Pe_r} \frac{ u(y) - \hat{U}_p}{D_y}\frac{\partial \mathcal{C}_p}{\partial \xi} = \frac{\partial^2 (D_y \mathcal{C}^{(1)})}{\partial y^2}.
%\end{equation}
%A solution of \eqref{eqn:CpC1} that satisfies the boundary condition \eqref{eqn:FP_nondim2} averaged over particle orientations is
\begin{equation}
    \label{eqn:C1_solution}
    \mathcal{C}^{(1)} = \mathrm{Pe_r} D^{-1}_y G(y)\frac{\partial \mathcal{C}_p}{\partial \xi}\qquad\text{where}\qquad G(y) = \int_{-1}^{y}\text{d}z\left\{ \int_{-1}^{z}D^{-1}_y\left(y'\right)\left( u(y') - u^{(0)}_m\right)\ \text{d}y'\right\}.
\end{equation}

At $\mathcal{O}(\varepsilon^3)$, equation \eqref{eqn:FP_nondim1} averaged over particle orientations gives
\begin{subequations}
\begin{equation}
\label{eqn:C2_equation}
D^{-1}_y\frac{\partial \mathcal{C}_m}{\partial T} = -\text{Pe}^2_r D^{-1}_y (u - u^{(0)}_m) G(y) \frac{\partial^2 \mathcal{C}_m}{\partial \xi^2} + D^{-1}_y \text{Pe}_r u^{(1)}_m \frac{\partial \mathcal{C}_m}{\partial \xi}\\ + 2\frac{\partial}{\partial y}\left(\frac{\langle D_{xy} g\rangle}{D_y}\right)\frac{\partial \mathcal{C}_m}{\partial \xi} + \frac{\partial^2}{\partial y^2}\left(D_y \mathcal{C}^{(2)}\right),
\end{equation}
where we have substituted equations \eqref{eqn:solutionC0} and \eqref{eqn:C1_solution} for $\mathcal{C}^{(0)}$ and $\mathcal{C}^{(1)}$, respectively. The boundary condition \eqref{eqn:FP_nondim2} averaged over particle orientations is 
\begin{equation}
 \label{eqn:C2_boundarycondition}
 \frac{1}{D_y}\langle g D_{xy}\rangle \frac{\partial \mathcal{C}_m}{\partial \xi} + \frac{\partial(D_y \mathcal{C}^{(2)})}{\partial y} = 0\qquad\text{at}\qquad y = \pm 1.
     \end{equation}
\end{subequations}
After taking the lateral average of equation \eqref{eqn:C2_equation}, using the boundary condition \eqref{eqn:C2_boundarycondition}, and choosing
\begin{equation}
u^{(1)}_m = -\frac{1}{2\mathrm{Pe_r}\overline{D^{-1}_y}}\left[D^{-1}_y\langle D_{xy} g\rangle\right]_{y = \pm 1}
\end{equation}
so as to again nullify the advective flux, we find
\begin{equation}
\label{eqn:Cp_t2_eqn}
\frac{\partial \mathcal{C}_m}{\partial T} = \kappa \mathrm{Pe}^2_r \frac{\partial ^2 \mathcal{C}_m}{\partial \xi^2}
\end{equation} where
\begin{equation}\label{final_kappa}
    \kappa = -\frac{\overline{D_y^{-1} G\left( u - u^{(0)}_m\right)}}{\overline{D^{-1}_y}}
\end{equation}
is the \emph{effective dispersion coefficient}. 

Finally, after returning to the laboratory frame $(x,t)$, we obtain
\begin{equation}\label{final_kappa_rods_52}
\frac{\partial \mathcal{C}_m}{\partial t}  +\mathrm{Pe}u_m\frac{\partial \mathcal{C}_m}{\partial x} = \kappa \mathrm{Pe}^2 \frac{\partial ^2 \mathcal{C}_m}{\partial x^2},
\end{equation}
where the mean speed particle speed, $u_m$, is  
\begin{equation}\label{final_um}
    u_m = \frac{\overline{D^{-1}_y u(y)}}{\overline{D^{-1}_y}} - \frac{1}{2 \mathrm{Pe}\overline{D^{-1}_y}}\left[D^{-1}_y\langle D_{xy} g\rangle\right]_{y = \pm 1}.
\end{equation}
Eq.(\ref{final_kappa_rods_52}) is the sought-after effective transport equation, analogous to equation \eqref{eqn:Taylor_result}, for ellipsoidal particles. We note that for spherical particles, where $p = D_y = 1$, we find that $u_m = 2/3$ and $\kappa = 8/945$, the latter consistent with equation \eqref{kappa_s}.} 

%\review{SJT: Remove? [[}In Figure \ref{fig:means}(a), we plot $u_m$ for $p \ge 1$ and various values of $\mathrm{Pe_r}$, from which we infer than $\mathcal{C}_p$ is dispersed relative to a frame travelling slower than the mean speed of the flow.\review{]]}

\begin{figure}
    \centering
    \includegraphics[width=1\textwidth]{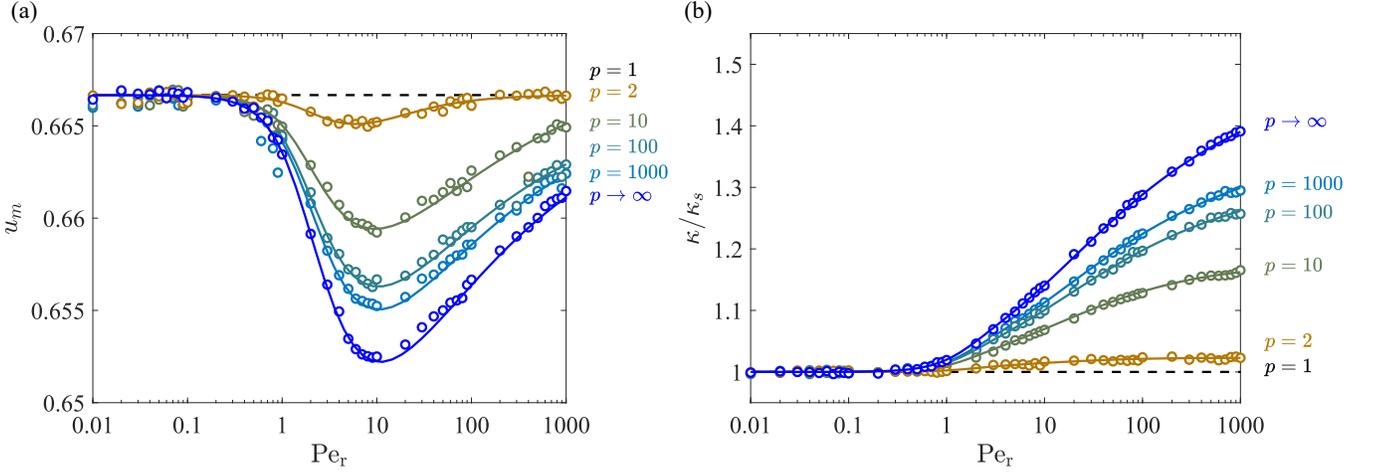}
    \caption{Plots of (a) the mean speed of the particles, $u_m$, and (b) the effective dispersion coefficient, $\kappa$, as a function of $\mathrm{Pe_r}$ for different aspect ratios, $p$ at $\mathrm{Pe}=1000$ . Circles are the results of our Monte Carlo simulations; solid lines are the theoretical predictions of $\kappa$ and $u_m$ given by equations \eqref{final_kappa} and \eqref{final_um}, respectively.}
    \label{fig:means}
\end{figure}

As shown in Figure \ref{fig:means}(a), even for elongated particles $(p > 1)$, the mean speed of the particles is approximately the mean speed of the flow ($u_m\approx 2/3$) when $\mathrm{Pe_r}\ll 1$. As $\mathrm{Pe_r}$ is increased, the particles migrate towards the channel walls where the local fluid velocity is smaller.  The different orientational distributions at each shear layer cause the particles to have different local $D_y$ values which is balanced by a net lateral migration velocity, as seen in Figure \ref{fig:drift_diffusion}.  There is a local minimum in the mean speed of the particles around $\mathrm{Pe_r} \approx 10$, as seen in Figure \ref{fig:means}. Beyond $\mathrm{Pe_r} \gtrsim 10$, the orientational distributions are quite similar at each shear layer away from the center of the channel making the local diffusion constant $D_y$ very similar across $y$. As a result, the overall lateral migration is actually smaller for large values of $\mathrm{Pe_r}$.
\par
\begin{figure}[ht]
    \centering
    \includegraphics[scale=0.65]{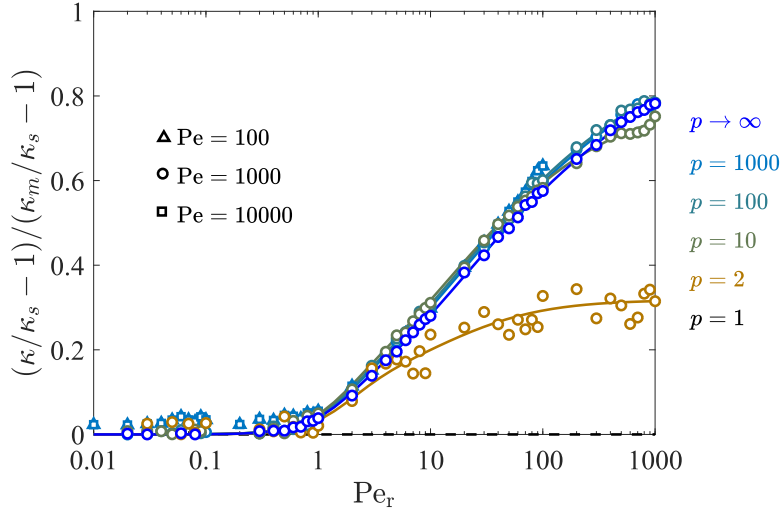}
    \caption{The fraction of the maximum possible dispersion enhancement achieved for a rod of aspect ratio $p$ as a function of the rotational Peclet number $\mathrm{Pe_r}$.  The data approximately collapses along a single curve for $p \gtrsim 10$.} 
    \label{fig:kappa theory scaling}
\end{figure}

Figure \ref{fig:means} demonstrates that the theoretical predictions and the Monte Carlo simulations show excellent agreement. Furthermore, in Figure \ref{fig:kappa theory scaling}, by normalizing the dispersion factor $\kappa$ with respect to its maximum possible value $\kappa_m$ and minimum possible value $\kappa_s$, the curves for different $p$ approximately collapse along one master curve. As $p$ decreases from approximately $10$ to $1$, the results diverge from the master curve and approach the flat line corresponding to Taylor's case of $p=1$. This observation suggests that in the limit of large $p$ and large $\mathrm{Pe}$, the asymptotic dispersion coefficient for elongated particles can be captured by a single curve, which depends only on $\mathrm{Pe_r}$.  This curve ultimately may serve as a simple and accessible correction factor to extend Taylor's result to the case of highly elongated rods.  

\review{The same asymptotic calculation can be readily to extended to the more general case when the rods are not confined to rotate strictly in the $xy$-plane, and is presented in Appendix B.  While the quantitative results differ, the tendency for the particles to align with the flow results in an enhanced dispersion via the same underlying physical mechanism.}

\section{Conclusion}
\label{conclusion}
 In this study, we have examined the bulk transport properties of elongated rods in a two-dimensional Poiseuille flow at high Peclet number using Monte Carlo simulation and semi-analytical theory inspired by Taylor's original work. For low rotational Peclet number, where rotational diffusion dominates rotational advection, the rods behave identically to spherical particles with similar values of the dispersion constant and mean particle speed. As the rotational Peclet number increases, \review{the shear-induced rotation starts dominating rotational diffusion and the rods align themselves more (on average) with the direction of the flow. This alignment effect makes it more difficult for the rods to diffuse across the streamlines as compared to spherical particles.} This reduced lateral diffusion directly results in an enhanced spreading of particles longitudinally, characterized by a larger value of the dispersion factor, as quantified by Monte Carlo simulations that in turn exhibit excellent agreement with our semi-analytical theory. Furthermore, the same theory allows us to characterize the mean speed of the particles, which always remains below the mean speed of the flow and exhibits a distinct minimum as the rotational Peclet number is varied. Our work reveals both when the non-spherical shape of the particle has an appreciable influence on the bulk dispersion properties as well as the conditions under which an elongated particle can be safely approximated as spherical (isotropic) in application.
\par
The present study focuses on two-dimensional flows but could be extended to three-dimensional parallel shear flows in future work.  While the quantitative details will inevitably differ, we similarly expect an enhanced spreading in three-dimensional flows due to the physical mechanism of flow alignment highlighted within the present work.  The subtle roles of channel geometry, more detailed particle shapes, and other more physically relevant boundary conditions on the dispersion process also deserve future attention. 

\appendix

\section{\review{Influence of rod orientation wall collision condition in Monte Carlo simulation}}
For all the previously presented simulation results, conservation of particles in the channel was ensured via a billiards-like reflection boundary condition wherein the orientation of the particles is unchanged following wall collision. \reviewtwo{For active Brownian particles, Peng and Brady similarly assumed that the orientation of the particles is unaffected by collisions with the walls of the channel \cite{peng2020upstream}. Similar to the billiards-like reflection condition, an alternative method to ensure the conservation of particles in the channel is the ``potential-free'' method where a suitably tuned force is applied to the particle only if it is predicted to escape the channel boundaries due to Brownian effects at a given time step \cite{heyes1993brownian}.  Both the billards-like reflection and ``potential-free'' methods are convenient idealizations to the detailed hydrodynamic boundary interactions, yet have been successfully used to model the no-flux boundary condition at the wall in prior works on Taylor dispersion \cite{aminian2016boundaries,peng2020upstream,wang2021vertical}.}  This section presents a discussion of two alternative idealized orientation collision conditions that affect the local alignment statistics and, consequently, the dispersion factor. The first case is when the rods are prescribed to align in the direction of the flow immediately after wall collision, which we will refer to as an aligning collision condition. The second case is when the rods have a uniformly random orientation following each wall collisions, which we will refer to as a randomizing collision condition.

\begin{figure}
    \centering
    \includegraphics[width=1\textwidth]{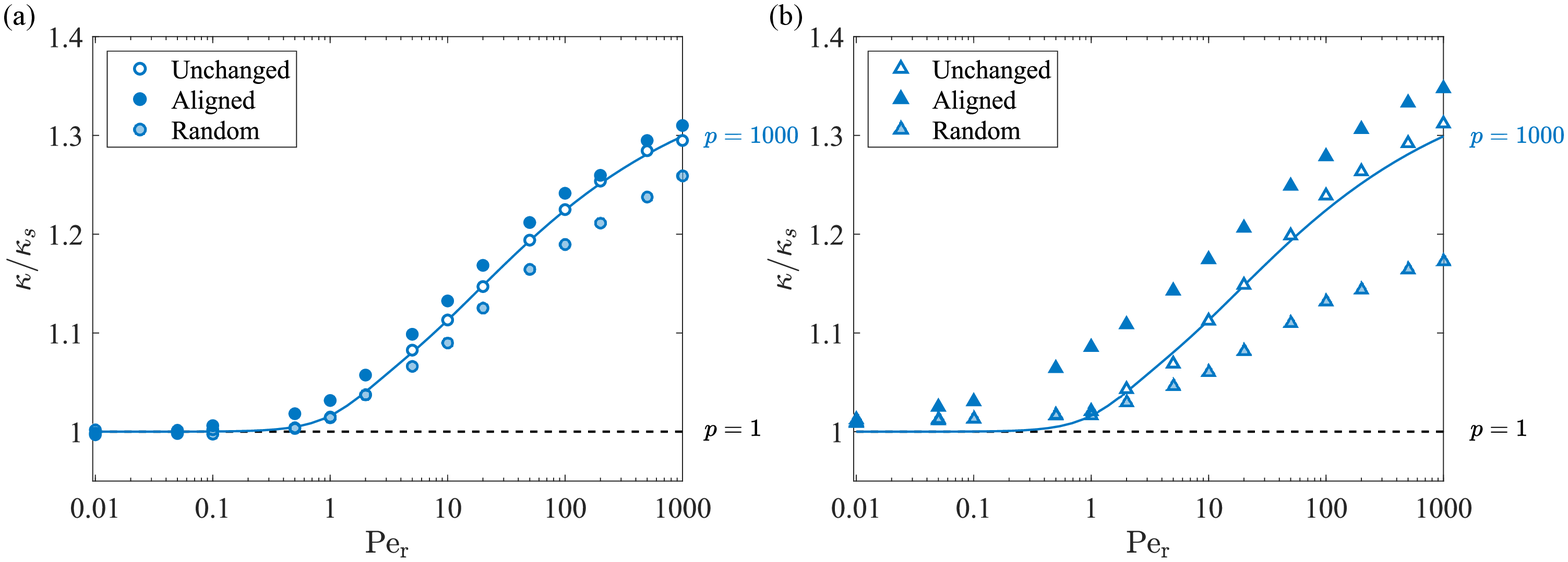}
    \caption{\review{Predictions for the effective dispersion coefficient assuming different boundary conditions in Monte Carlo simulations with (a) $\mathrm{Pe}=1000$ and (b) $\mathrm{Pe}=100$. The unfilled points represents the case when the orientation of the particles is unaffected by the collisions, the filled points represent the case when the rods align themselves in the direction of the flow after collision and, the shaded points refer to the case when the orientation of the particles is fully randomized after each collision.  The solid lines indicates the theoretical prediction derived in \S\ref{sec:dispCoeff_meanPS} (equation (\ref{eqn:Cp_t2_eqn})).}}
    \label{fig:Appendix_A_MC_BC}
\end{figure}

As demonstrated in Figure \ref{fig:Appendix_A_MC_BC}(a), for the case of aligning collisions, the particles' overall alignment with the flow is stronger which results in greater dispersion.  In contrast, randomizing collisions systematically reduces dispersion by weakening overall alignment.  The unchanged wall condition sits between these extremes, and is best predicted by the continuum theory presented in \S\ref{sec:dispCoeff_meanPS}.  We repeated the same set of simulations for a lower value of $\mathrm{Pe}$ in Figure \ref{fig:Appendix_A_MC_BC}(b) and observed the same overall trends, but with an increased deviation between the predictions from the three idealized boundary conditions. One way to interpreter this finding is as follows: for a fixed $\mathrm{Pe_r}$ the dispersion is decreasingly sensitive to the details of particle-wall interactions as $\mathrm{Pe}$ is increased.  For the physically relevant regime defined by $\mathrm{Pe}\gg \mathrm{Pe_r}$ (corresponding to $a_p \ll a$: equation (\ref{particle channel})), the timescale for equilibration of the orientational dynamics is much faster than the translational timescales of the problem.  Thus following a collision, particles orientations rapidly relax to their steady-state orientational distributions.  Consistent with this interpretation, and as evidenced in these Monte Carlo simulations, the overall dispersion statistics are most weakly influenced by the details of the wall collisions when $\mathrm{Pe}\gg \mathrm{Pe_r}$. 

These results ultimately highlight the role of the assumed particle-wall dynamics on the long-term dispersion behavior.  Considering the detailed hydrodynamics associated with particle-wall collisions would thus inevitably affect the overall spreading statistics, and should be explored in future work.  

% In the case of random collisions, the particles' overall alignment along the direction of flow is weaker, and we see a lower dispersion factor. Finally, the figure also shows that the simulations where the orientations are unaffected by the collisions show the best agreement with the theoretical estimation. In reality, it is hard to determine the exact type of collisions the rods experience as there could be a combination of the above discussed simple cases or some more complicated multiple collisions.

\section{\review{Unconstrained rotation: 3D infinite parallel plates}}
\review{
In this section, we extend the analytical prediction based on the continuum model to the three-dimensional case of infinite parallel plates, where the rods have two degrees of rotational freedom.  For this calculation, we assume there are no gradients along the $z$ direction (into the page in relation to Figure \ref{fig:coordinates}). The diffusion tensor $\mathsf{D}$ for the governing Fokker-Planck equation for particles in 3D is given by \cite{brenner1974transport} 
\begin{equation}
    \mathsf{{D}}(\theta,\phi) = \mathbf{e} \ \mathbf{e} D_{\parallel} + (\mathsf{{I}} - \mathbf{e} \ \mathbf{e})D_{\perp},
\end{equation}
where $\mathbf{e}=\cos\theta \cos \phi \ e_x + \sin\theta \cos \phi \ e_y + \sin \phi \ e_z.$ In the present work, $\theta$ is the angle the rod makes along the $xy$ plane (with $\theta=0$ corresponding to the positive $x$-axis) and $\phi$ is the angle made by the rod along the $xz$ plane (with $\phi=0$ corresponding to the positive $x$-axis). In the $xyz$ (laboratory) basis, the components of the translational diffusion tensor are, 
%\begin{equation}\label{egn:Difften3d}
%    \begin{bmatrix} D_{xx}(\theta,\phi) & D_{xy} (\theta,\phi) & D_{xz} (\theta,\phi) \\ D_{xy} (\theta,\phi) & D_{yy} (\theta,\phi) & D_{yz} (\theta,\phi) \\ D_{xz} (\theta ,\phi) & D_{yz}(\theta,\phi) & D_{zz} (\theta,\phi) \end{bmatrix} = \begin{bmatrix} D_{\parallel} \cos^2 \theta \cos^2 \phi + D_{\perp} ( 1 - \cos^2 {\theta} \cos^2{\phi}) & (D_{\parallel} \cos \theta \sin \theta - D_{\perp} \cos \theta \sin \theta ) \cos^2 \phi \\ (D_{\parallel} \cos \theta \sin \theta - D_{\perp} \cos \theta \sin \theta ) \cos^2 \phi & D_{\parallel} \sin^2 \theta \cos^2 \phi + D_{\perp} (1 - \sin^2 \theta \cos^2 \phi) \end{bmatrix}.
%\end{equation}
\begin{subequations}
\begin{equation}\label{egn:Difften3d}
    \begin{bmatrix} D_{xx}(\theta,\phi) & D_{xy} (\theta,\phi) & D_{xz} (\theta,\phi) \\ D_{xy} (\theta,\phi) & D_{yy} (\theta,\phi) & D_{yz} (\theta,\phi) \\ D_{xz} (\theta ,\phi) & D_{yz}(\theta,\phi) & D_{zz} (\theta,\phi)  \end{bmatrix}
\end{equation}
where
\begin{eqnarray}
    D_{xx} (\theta,\phi) &=&  D_{\parallel} \cos ^2(\theta ) \cos ^2(\phi )+D_{\perp} \left(1-\cos ^2(\theta ) \cos ^2(\phi )\right), \\ D_{xy} (\theta,\phi) &=& D_{\parallel} \sin (\theta ) \cos (\theta ) \cos ^2(\phi  )-D_{\perp} \sin (\theta ) \cos (\theta ) \cos ^2(\phi ), \\ D_{xz} (\theta, \phi) &=& D_{\parallel} \cos (\theta ) \sin (\phi ) \cos (\phi )-D_{\perp} \cos (\theta ) \sin (\phi ) \cos(\phi ), \\ D_{yy}(\theta,\phi) &=& D_{\parallel} \sin ^2(\theta ) \cos ^2(\phi)+D_{\perp} \left(1-\sin ^2(\theta ) \cos ^2(\phi )\right), \\ D_{yz} (\theta,\phi) &=& D_{\parallel} \sin (\theta ) \sin (\phi ) \cos (\phi )-D_{\perp} \sin (\theta ) \sin (\phi ) \cos(\phi ), \\ D_{zz} (\theta,\phi) &=&  D_{\parallel} \sin ^2(\phi )+D_{\perp} \left(1-\sin ^2(\phi )\right).
\end{eqnarray}
\end{subequations}
%\begin{equation}
%\begin{bmatrix}
 %D_{\parallel} \cos ^2(\theta ) \cos ^2(\phi )+D_{\perp} \left(1-\cos ^2(\theta ) \cos ^2(\phi )\right) & D_{\parallel} \sin (\theta ) \cos (\theta ) \cos ^2(\phi
  % )-D_{\perp} \sin (\theta ) \cos (\theta ) \cos ^2(\phi ) & D_{\parallel} \cos (\theta ) \sin (\phi ) \cos (\phi )-D_{\perp} \cos (\theta ) \sin (\phi ) \cos
   %(\phi ) \\
 %D_{\parallel} \sin (\theta ) \cos (\theta ) \cos ^2(\phi )-D_{\perp} \sin (\theta ) \cos (\theta ) \cos ^2(\phi ) & D_{\parallel} \sin ^2(\theta ) \cos ^2(\phi
  % )+D_{\perp} \left(1-\sin ^2(\theta ) \cos ^2(\phi )\right) & D_{\parallel} \sin (\theta ) \sin (\phi ) \cos (\phi )-D_{\perp} \sin (\theta ) \sin (\phi ) \cos
   %(\phi ) \\
 %D_{\parallel} \cos (\theta ) \sin (\phi ) \cos (\phi )-D_{\perp} \cos (\theta ) \sin (\phi ) \cos (\phi ) & D_{\parallel} \sin (\theta ) \sin (\phi ) \cos (\phi
  % )-D_{\perp} \sin (\theta ) \sin (\phi ) \cos (\phi ) & D_{\parallel} \sin ^2(\phi )+D_{\perp} \left(1-\sin ^2(\phi )\right) \\
%\end{bmatrix}.
%\end{equation}
The $xy$ components of this tensor are identical to equation (\ref{difftensor2}) when $\phi=0$, which corresponds to the constrained problem considered hitherto. For the 3D case, we define the orientationally averaged diffusivity as \begin{equation}
    \bar{D} = \frac{D_{\parallel} + 2 D_{\perp}}{3}.
\end{equation}
The dimensional form of the conservation equation for the probability distribution $P(\bold{x},\theta,\phi,t)$, for the particles is
\begin{subequations}\label{3d_dimensional}
\begin{equation}
\begin{aligned}
    \frac{\partial P}{\partial t} = -u(y) \frac{\partial P}{\partial x} + D_{xx}(\theta,\phi) \frac{\partial^2 P}{\partial x^2} + 2 D_{xy} (\theta, \phi) \frac{\partial^2 P}{\partial x \partial y} + 2 D_{xz} (\theta,\phi) \frac{\partial^2 P}{\partial x \partial z} + 2 D_{yz} (\theta, \phi) \frac{\partial^2 P}{\partial y \partial z}  D_{yy}(\theta,\phi) \frac{\partial^2 P}{\partial y^2} \\ + D_{zz} (\theta,\phi) \frac{\partial^2 P}{\partial z^2} +  D_{\theta} \left[\frac{1}{\cos^2 {\phi}} \frac{\partial^2 P}{\partial \theta^2} + \frac{1}{\cos \phi} \frac{\partial}{\partial \phi} \left( \cos \phi \frac{\partial P}{\partial \phi} \right) \right] - \left[\frac{\partial}{\partial\theta}\left(\omega_{\theta} g\right) + \frac{1}{\cos\phi}\frac{\partial}{\partial\phi}\left(\cos\phi \ \omega_{\phi} g\right)\right]
\end{aligned}
\end{equation}
where 
\begin{equation}
    \omega_{\theta}(\theta) = \frac{\Dot{\gamma}(y)}{2}(1 - \beta\cos 2\theta),\qquad \omega_{\phi}(\theta,\phi) = \frac{\Dot{\gamma}(y)}{4}\beta\sin 2\theta\sin 2\phi,\qquad\text{and}\qquad \beta = \frac{p^2 - 1}{p^2 + 1}.
\end{equation}
The symmetry of rod-shaped particles makes the the probability distribution periodic in $\theta$ and $\phi$, with $P(\bold{x},\theta + \pi,\phi,t) = P(\bold{x},\theta,\phi,t)$ and $P(\bold{x},\theta,\phi+\pi,t) = P(\bold{x},\theta,\phi,t)$. We also demand the no-flux boundary condition at the walls, 
\begin{equation}
    (\bold{J} \cdot \hat{y}) = D_{xy} (\theta,\phi) \frac{\partial P}{\partial x} + D_{yz} (\theta,\phi) \frac{\partial P}{\partial z} + D_{yy}(\theta,\phi) \frac{\partial P}{\partial y} = 0 \qquad\text{at}\qquad y = \pm a.
\end{equation}
\end{subequations}
Upon non-dimensionalizing in the same way as Section \ref{sec:FP}, moving into the mean frame of reference of the particles, and employing the assumption of no gradients in the $z$ direction, the conservation equation becomes 
\begin{subequations}
\label{eqn:FP_nondims3d}
 \begin{equation}
     \label{eqn:FP_nondim3d1}
     \varepsilon \frac{\partial P}{\partial t} = -\varepsilon \mathrm{Pe_r}  \left(u(y) - u_m \right)\frac{\partial P}{\partial X} + \varepsilon^3 D_{xx}(\theta,\phi)\frac{\partial^2 P}{\partial X^2} + 2\varepsilon^2 D_{xy}(\theta,\phi)\frac{\partial ^2 P}{\partial X\partial y} + \varepsilon D_{yy}(\theta,\phi)\frac{\partial^2 P}{\partial y^2} + \mathcal{L} P(\theta, \phi ; y),
     \end{equation}
     with zero-flux boundary condition
     \begin{equation}
     \label{eqn:FP_nondim3d2}
     \varepsilon D_{xy} (\theta,\phi)\frac{\partial P}{\partial X } + D_{yy} (\theta,\phi) \frac{\partial P}{\partial y} = 0\qquad\text{at}\qquad y = \pm 1,
 \end{equation}
\end{subequations}
where
\begin{equation}
\begin{aligned}
    \mathcal{L} P(\theta, \phi ; y)  =  \left[\frac{1}{\cos^2 {\phi}} \frac{\partial^2 P}{\partial \theta^2} + \frac{1}{\cos \phi} \frac{\partial}{\partial \phi} \left( \cos \phi \frac{\partial P}{\partial \phi} \right) \right] \\ +2 y \mathrm{Pe_r} \left[\frac{\partial}{\partial \theta} (\omega_{\theta} \  P) +   \frac{1}{\cos \phi} \frac{\partial}{\partial \phi} \left( \cos \phi \  \omega_{\phi}  \ P \right)\right].
\end{aligned}
\end{equation}
Using the same asymptotic procedure as outlined in Section \ref{sec:dispCoeff_meanPS}, at leading order we obtain
\begin{subequations}
\label{eqn:g3D}
\begin{equation}
\label{eqn:g3D_eqn}
\left[\frac{1}{\cos^2\phi}\frac{\partial^2 g}{\partial\theta^2} + \frac{1}{\cos\phi}\frac{\partial}{\partial\phi}\left(\cos\phi\frac{\partial g}{\partial\phi}\right)\right] + 2y \mathrm{Pe_r}\left[\frac{\partial}{\partial\theta}\left(\omega_{\theta} g\right) + \frac{1}{\cos\phi}\frac{\partial}{\partial\phi}\left(\cos\phi \ \omega_{\phi} g\right)\right] = 0
\end{equation}
subject to the normalization condition
\begin{equation}
\int_0^{2\pi}\int_0^{\pi} g\cos\phi\ \text{d}\theta\ \text{d}\phi = 1,
\end{equation}
\end{subequations}
where
\begin{equation}
\omega_{\theta}(\theta) = \frac{1}{2}(1 - \beta\cos 2\theta),\qquad \omega_{\phi}(\theta,\phi) = \frac{1}{4}\beta\sin 2\theta\sin 2\phi,\qquad\text{and}\qquad \beta = \frac{p^2 - 1}{p^2 + 1}.
\end{equation}
Following \cite{nitsche1997shear}, the boundary value problem \eqref{eqn:g3D} was solved using a truncated generalized Fourier (Laplace) series of the form
\begin{equation}
    \label{eqn:genFS}
    g = \frac{1}{4\pi} + \sum_{l = 1}^{M}A_l(y) N_{2l}(\sin\phi) + \sum_{m = 1}^{M}\sum_{l = m}^{M}\left[B_l^m(y) N_{2l}^{2m}(\sin\phi)\cos(2m\theta) + C_l^m(y) N_{2l}^{2m}(\sin\phi)\sin(2m\theta)\right],
\end{equation}
where $N_l^m$ are the fully normalized associated Legendre functions \cite{abramowitz1988handbook}, related to the unnormalized associated Legendre functions, $P_l^m$, by 
\begin{equation}
N^m_l = (-1)^m\sqrt{\frac{\left(l + \frac{1}{2}\right)(l-m)!}{(l + m)!}}P^m_l.
\end{equation}
We note that symmetry of particle orientations under $(\theta,\phi)\rightarrow(\theta + \pi,\phi + \pi)$ eliminates both even degrees and orders of the Legendre functions. Furthermore, owing to the form of the rotation rates $\omega_\theta$ and $\omega_\phi$, we may restrict our attention to the domain $0 \le \theta\le \pi$ and $0\le\phi\le\pi/2$. Following an analogous procedure to that outlined in \S\ref{sec:dispCoeff_meanPS}, inserting the expansion \eqref{eqn:genFS} into \eqref{eqn:g3D_eqn} and enforcing the differential equation at every point
\begin{equation}
\theta_i = \frac{\pi i}{I},\qquad i = 1,\ldots,I\qquad\text{and}\qquad \phi_j = \frac{\pi j}{2 J},\qquad j = 1,\ldots,J,
\end{equation}
results in an overdetermined system of equations of dimension $IJ\times M(M + 2)$ for the coefficients $A_l(y)$, $B_l^m(y)$, and $C_l^m(y)$. For each value of $y$ (discretized from $y=0$ to $y=1$ using 200 equally spaced values), the resulting system was again solved using a standard QR least-squares algorithm in MATLAB with $I = 72$, $J = 144$, and $M = 32$. For $p=2$, we only needed $M=16$ modes for convergence.

\review{Having now solved the leading order (orientational) problem, solving the higher order equations becomes identical to the procedure outlined in \S\ref{sec:dispCoeff_meanPS}.  The final expressions for $\kappa$ and $u_m$ are also the same (equations (\ref{final_kappa}) and (\ref{final_um}), respectively), but with orientational averages now computed over both angles $\theta$ and $\phi$, specifically:
\begin{equation}
%D_y= \langle D_{yy} g\rangle \qquad\text{and}\qquad 
\langle \Box \rangle = \int_0^{2\pi}\int_0^{\pi} \Box \cos\phi\ \text{d}\theta\ \text{d}\phi. 
\end{equation}

 \begin{figure}
     \centering
     \includegraphics[width=1\textwidth]{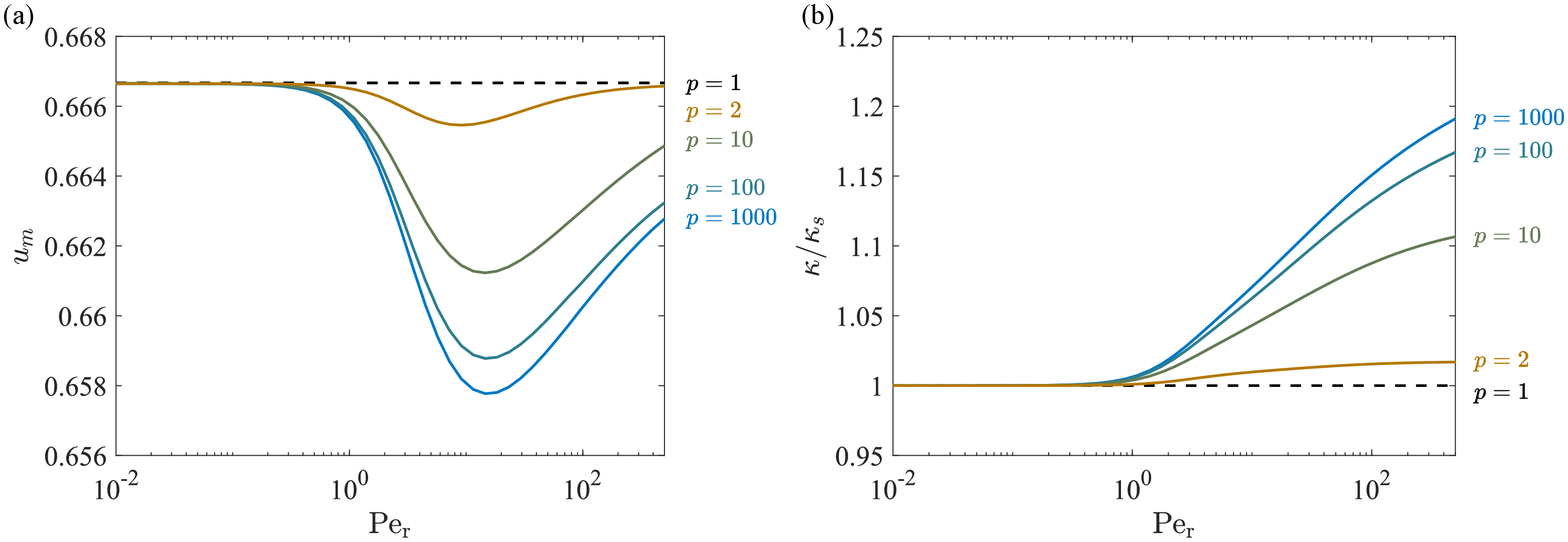}
         \caption{\review{ Theoretical predictions for mean particle speed and dispersion factor for the case of {\it unconstrained} rotation at $\mathrm{Pe}=1000$, as described in Appendix B. Plots of (a) the mean speed of the particles, $u_m$ and (b) the effective dispersion factor, $\kappa$, as a function of $\mathrm{Pe_r}$ for different aspect ratios $p$.}}
     \label{fig:Appendix_B}
 \end{figure}
 
Predictions for the mean particle speed, $u_m$, and dispersion factor, $\kappa$, are presented in Figure \ref{fig:Appendix_B}.  The overall trends are remarkably similar to the constrained rotation problem considered in the main text (Figure \ref{fig:means}), but with the departures from the spherical case reduced in magnitude.

}}

\begin{acknowledgments} 
We acknowledge funding received from NSF through award CMMI-1634552 and from the Brown OVPR Salomon Research Award. The Monte Carlo simulations were conducted using computational resources and services at the Center for Computation and Visualization, Brown University. The authors would like to thank Qian Chen and Brandon Vorrius %from the Warren Alpert Medical School at Brown University 
for fruitful discussions that helped conceive this research \review{and Francesca Bernardi and Manuchehr Aminian for useful suggestions and feedback.}% question. 
%Prof. Qian Chen and Brandon Vorrius from the Warren Alpert Medical School at Brown University for fruitful discussions that helped conceive this research question. 
\end{acknowledgments}

\bibliography{references}

\end{document}